\documentclass[twocolumn]{aastex631}

\usepackage{colortbl}
\usepackage{xcolor}
\usepackage{array}
\usepackage{CJK}

\shorttitle{Closely Packed Multiple Planetary Systems}
\shortauthors{Wang \& Lin}

\graphicspath{{./}{figures/}}

\begin{document}
\begin{CJK*}{UTF8}{gbsn}

\title{Dynamical Evolution of Closely Packed Multiple Planetary Systems Subject to Atmospheric Mass-Loss}

\correspondingauthor{Su Wang}
\email{wangsu@pmo.ac.cn}

\author{Su Wang (王素)}
\affiliation{CAS Key Laboratory of Planetary Sciences, Purple Mountain Observatory\\
Chinese Academy of Sciences,
Nanjing 210008, China }
\affiliation{CAS Center for Excellence in Comparative Planetology, Hefei 230026, China }

\author{D.N.C. Lin (林潮)}
\affiliation{Department of Astronomy and Astrophysics \\
University of California, Santa Cruz, CA 95064, USA}
\affiliation{Institute for Advanced Studies \\
Tsinghua University, Beijing 100086, China}

\begin{abstract}

A gap in exoplanets' radius distribution has been widely
attributed to the photo-evaporation threshold of their
progenitors' gaseous envelope. Giant impacts can also 
lead to substantial mass-loss. The outflowing
gas endures tidal torque from the planets 
and their host stars. Alongside the planet-star
tidal and magnetic interaction, this effect leads to 
planets' orbital evolution. In multiple super-Earth systems, especially in those which are closely
spaced and/or contain planets locked in mean motion
resonances (MMRs), modest mass-loss can lead to 
dynamical instabilities.  In order to place some 
constraints on the extent of planets' mass-loss,  
we study the evolution of a series of idealized systems of multiple 
planets with equal masses and a general scaled separation. 
We consider mass-loss from one or more planets either in 
the conservative limit or with angular momentum loss from 
the system. We show that the stable preservation of idealized 
multiple planetary systems requires either a wide initial separation
or a modest upper limit in the amount of mass-loss. This 
constraint is stringent for the multiple planetary systems in 
compact and resonant chains.
Perturbation due to either impulsive giant impacts between super-Earths or greater than a few percent mass-loss can lead to dynamical 
instabilities.

\end{abstract}


\section{Introduction} 
More than 5000 explanets around mature stars have been identified and 
confirmed \citep{Howard2012, Bata13, Fressin2013, petigura2013, Fab14, Silburt2015, Zink2019, kunimoto2020}.  
In order to utilize their dynamical properties to extract clues and place
constraints on scenarios of planet formation, we must take into consideration
planets' post orbital formation and structural 
evolution \citep{Ida2010, Ida2013, Jin14, helled2014}.  
In this quest, over 800 members of multiple planetary 
systems provide valuable clues and best tests for theoretical models \citep{lissauer2011, lovis2011, Fab14,  Weiss2018,Millholland2021}.
 
The locations of innermost planets in multiple super-Earth systems are often 
in the proximity of their host stars.  These short-period planets
are exposed to the intense flux of X-ray, FUV, and UV photons,
especially during their host stars' infancy \citep{Baraffe2005, Murray2009, Owen2012, Erkaev2016, Kubyshkina2018, Sai18, Inonov18, King2021}.
Photoionization of H/He-dominant planetary envelopes can lead to
a substantial amount of mass-loss from low-mass close-in planets,
while only a small fraction of the gaseous envelope may be removed
from the more massive and distant planets \citep{Inonov18}.
Theoretical analyses suggest the evaporation rate is a sensitive
function of planets' core and envelope masses. An estimation on the
evaporation threshold leads to the prediction of a mass-loss
dichotomy and a bimodal distribution in the apparent sizes among
planets over a small range of masses \citep{Wu13, Owen2013, Owen17, Lopez2013}.

This theoretical prediction has been confirmed by an observed gap
(based on Kepler's data) in the planets' size distribution between
$\sim 1.3-2 R_\oplus$ over a range of periods \citep{Fulton17, Petigura17, Ber18, Fulton2018,  Berger2020, Van2018, Petigura2022}.
The masses of planets on either side of the gap have been determined
from the transit timing variation (TTV) \citep{Lithwick2012, Hadden2014, Hadden2017, Wu13, Jon16} and
radial velocity measurements \citep{Weiss14, Hadden2014, Dressing2015, Wolfgang2016, Chen2017}.  Their nearly uniform
masses match well with the theoretically predicted threshold
transition across an ``evaporation valley'' \citep{Lopez2013, Owen2013, Owen17, Weiss2013, Wu13, Jin14, OwenA2019, Mordasini2020}.

The success of the theoretical prediction introduces a new
conundrum concerning the progenitors of the super-Earths on
either side of the period gap.  If the main cause of the gap
is due to a critical condition for photo-evaporation, the
progenitors of all planets on either side of the gap must
have formed with nearly identical core masses ($M_c
\sim 10 M_\oplus$) and a modest fraction ($f \sim$ a
few $\%$) of envelope masses \citep{ Madhusudhan2019, Owen2019, Wu2019, Zhang2020, Rogers2021, Rogers2023}. 

The near uniformity of $M_c$ provides both clues and constraints on
the core accretion scenario \citep{idalin04} which assumes an initial
emergence of planetesimals through runaway growth \citep{Dullemond2005, Kenyon2009, 
 Garaud2013}, gravitational instability \citep{Goldreich1973, 
Weidenschilling1993, Youdin2002, GaraudLin2007}, streaming instability (SI) 
\citep{Youdin2005b} or vortices trapping \citep{Johansen2007} of 
small dust in protostellar disks.  Thereafter, these building blocks rapidly 
coagulate until a few oligarchic embryos emerge (with masses from a fraction to a 
few $M_\oplus$) which consumes all of the residual planetesimals in their feeding zones 
\citep{KokuboIda1998}. However, embryos' growth can be further supplied by fast migrating 
optimum-sized grains (i.e., approximately millimeter- to meter- sized pebbles; 
\citep{OrmelKlahr2010}), until they attain the pebble-isolation mass, typically 
a few to 10$M_\oplus$ and evolve into protoplanetary cores. Subsequently, cores 
induce local maxima in the surface density and pressure distribution for the disk 
gas which suppresses their growth \citep{LambrechtsJohansen2012, Lambrechtsetal2014, 
Ormel2017, Bitschetal2018}. This effect quenches the cores’ growth via further 
pebble accretion. The gas accretion rate is regulated by both opacity and entropy 
advection \citep{Ikomaetal2000, Rafikov2006, Leeetal2014, PisoYoudin2014, Pisoetal2015,
LeeChiang2015, LeeChiang2016}. 
The accumulation of pebbles near their tidal barrier also increases 
their collision frequency with fragmentary byproducts in the form
of sub-mm grains which are well coupled to and dragged by the 
gas flow. This process elevates the effective dust opacity, reduces the 
radiative flux in the envelope, limits the envelope's asymptotic  
masses ($M_e$) through gas accretion, and preserves super-Earths in 
the stellar proximity at $\sim 0.1$ AU \citep{Chenetal2020}.

\begin{figure*}[htp]
\begin{center}
  \epsscale{1.0}
  \plotone{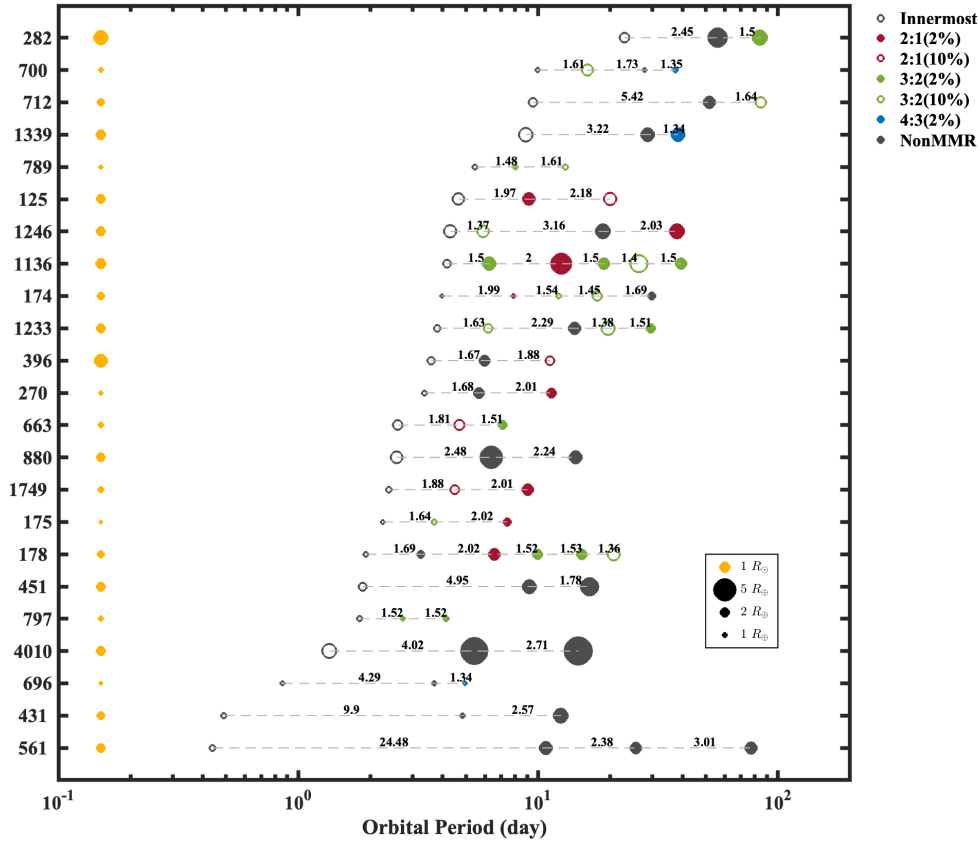}
 \caption{The distribution of planetary systems which contain more than three planets observed by TESS mission. The grey circles mean the innermost planet in the system, and the inner one in a planet pair is labeled in red if the planet pair is in or near 2:1 MMR, red sold dots represent the period ratio deviations of 2.0 by no more than 2\%, while the red circles represent the period ratio deviations of 2.0 by no more than 10\%. The green sold dots and circles show the period ratio results of 3:2 MMR, and the blue sold dots show the results of 4:3 MMR. The grey sold dots mean no obvious sign of first order MMR relationship between two adjacent planets.
 \label{multi}}
 \end{center}
\end{figure*}

These considerations provide a natural explanation for the nearly uniformity in
$M_c$ of close-in super-Earths in the context of {\it in-situ} formation
scenario \citep{ChatterjeeTan2014, Liu17b}.  However, it is not clear whether 
there may be a dispersion in the metallicity and mass of their envelopes ($M_e$),  
especially in multiple planetary systems where the gas and grain supplies may
vary among members with a range of semi-major axis.  Moreover, some multiple 
systems are locked in mean motion resonances (MMRs), including  chain configurations
which suggest the likelihood of their orbital migration \citep{Gi17, Leleu2021}.  Similar to hot Jupiters
\citep{lin96}, some super-Earths may have formed at a few AU from their host 
stars and migrated over large distances. At these distances, the cores' critical 
masses for inducing pebble isolation and suppressing gas accretion may differ
from those close to their host stars \citep{Chenetal2020}. Such diversity
may also introduce dispersion in the asymptotic values of both  $M_c$ 
and $M_e$ among infant super-Earths.

Rapid convergent migration can also lead to orbit crossing and giant impacts.
Even if super-Earths' progenitors had a substantial gaseous envelope, such a 
population could be the byproducts of giant impacts \citep{Liusf15} during the
depletion of their natal disks. These events can lead to increases in their 
$M_c$ and losses in their envelope masses.

In this paper, we consider a potential implication of
substantial mass-loss from super-Earths' gaseous envelope.
This possibility would arise naturally if some super-Earths formed with predominantly gaseous envelopes
or contained mostly volatile ices and subsequently
relocated to the proximity of their host stars to
endure intense stellar irradiation.  In principle,
photo-evaporation would remove a large fraction of their
initial mass in their envelope if their core mass was
below some threshold value for atmospheric retention.
But, the outflowing gas is subjected to the tidal
torque and the wind from their host stars 
\citep{Carroll-Nellenback2017} which induce angular
momentum transfer between the evaporated gas and
the orbits of the planets. This interaction leads
to fractional changes in the semi-major axes of
the planets through an amount of mass they have lost. With a set of idealized models,
we first demonstrate (in \S2) that this effect
can lead to dynamical instabilities in closely packed
multiple planetary systems.  For computational simplicity,
we construct these idealized models with a set uniform
planetary masses, scaled locations and separation (in units of
Roche radii).

Although idealized models are useful to illustrate the
destabilization tendency induced by the mass-loss, they
need to be generalized to more realistic multiple planetary
systems with a wide range of observed kinematic configurations.
The list of known multiple planetary systems is likely to grow
rapidly with the ongoing confirmation campaign for hundreds
of planetary candidates in the Kepler dataset \citep{Bata13,
Mazeh13, Fab14, Dr17}. In addition, TESS mission has
already found several multi-planet systems and many
more are anticipated \citep{quinn19}, as shown in Figure \ref{multi} which includs candidate and confirmed systems. In some of the closely
packed systems, modest orbital migration induced by the
loss of envelope's mass or impulsive energy and angular
momentum changes due to giant impacts may introduce
non-negligible perturbation and disturb the stability
of these systems especially if they are closely packed,
similar to the TRAPPIST-1 system \citep{Gi17}, or TOI-178 system \citep{Leleu2021}.

Some compact multiple planetary systems reside in nearly commensurable
orbits with their members' period ratios close to some low-order
integer fractions. Presumably, these systems have maintained
their present-day configurations over the age of their host stars.
Nevertheless, the resonant planets exchange orbital energy and
angular momentum which lead to variations in their semi-major
axes, eccentricities, and libration in their relative orbital
phases. In \S3, we scout the island of stability of compact systems
with multiple MMRs.  Based on a migration model, we robustly reproduce
a series of multiple planetary systems with MMRs. With these initial
orbits, we show that these MMR configurations occupy narrowly confined
islands of stability. We impose small impulsive changes in the spacing
between adjacent planets in these systems and show that they can lead to
rapid evolution in their orbital elements, including transitions to
dynamical instability. We then introduce protracted mass-loss and
orbital evolution to explore some quantitative constraints on the
amount and rate of photo-evaporation based on its consequential
change to the stability of the systems.  Finally, we summarize our
results and discuss their implications in \S4.

\section{Stability of Systems with Mass-Loss of the Innermost Planets}

In this section, we illustrate the consequence of mass-loss
on an idealized system of multiple planets with equal initial
mass and scaled spacing, in units of their local Roche radius.
This set of initial conditions are designed to highlight the
destabilizing effect of mass-loss on compact multiple planetary
systems. We assume envelope is lost after the depletion of these
planets' natal disk and introduce a simple prescription to
approximate its influence on the orbit of the mass-losing planet.

\subsection{The timescales of mass-loss and orbital migration}
\label{sec:massloss}
Herein, we approximate the loss of the planet's envelope proceeds
on a constant, characteristic time scale $\tau_m$ such that
\begin{equation}
\tau_m=-\frac{m_p}{\dot m_p},
\end{equation}
where $\dot m_p$ is the mass-loss rate of planet.
The planet's orbital angular momentum $L$ is
\begin{equation}
L=m_p\sqrt{GM_*a_p (1-e^2)},
\end{equation}
such that
\begin{eqnarray}
\frac{\dot a_p}{a_p}=2 \left(
\frac{\dot L}{L}-\frac{\dot m_p}{m_p} +\frac{e {\dot e}}{(1-e^2)} \right).~~~~
\end{eqnarray}

If the evaporation of the planet proceeds after they are
tidally synchronized, the planet's internal tidal dissipation
may also be efficiently intense to render a non negligible ${\dot e}$.  In this case,
photo-ionization occurs mostly on the day side facing the
host star.  Around quiet mature stars (those without an
intense stellar wind or/and strong magnetic field), gas
leaves the planets' envelope via their inner Lagrangian point
and mostly falls toward the star \citep{Carroll-Nellenback2017}.  In this case, angular momentum
of the systems is essentially conserved (i.e. $\dot L=0$) and
the planets' orbits expand on a time scale $\tau_a=a_p/
\dot a_p=\tau_m/2$.

Planet-host stars' UV flux is most intense during their
infancy when they also blow powerful wind which may exert strong
ram pressure on the evaporated gas from the planets' surface.
Planetary magnetic field may also modify the outflow direction
of the evaporated gas \citep{Bisikalo2013,ZhilkinBisikalo2019}.  Under some circumstances, angular
momentum may be lost from the systems \citep{Carroll-Nellenback2017},
while the planets' orbital
decay on a similar time scale $\tau_a$ (Tang et al in preparation).

In order to take into account of these diverse possible outcomes,
we introduce a simple idealized prescription to characterize planets' mass-loss with
\begin{equation}
m_p (t)=m_0[1+{\rm A}exp(-t/\tau_m)]/(1+{\rm A}) .
\end{equation}
Main model parameters are: 1) the planets' initial mass ($m_0$), 2) the fractional mass-loss amplitude $A$, and
3) the mass-loss time scale $\tau_m$. The model parameters used here are chosen for illustrative purposes and they can easily be scaled to the appropriate values applicable to real observed systems.  We approximate the planets'
orbital evolution with
\begin{equation}
{\dot a \over a} = {2 f_p \dot m_p \over m_p}
={- 2 f_p \over \tau_m}
\left( {{\rm A exp}(-t/\tau_m) \over
1+{\rm A exp}(-t/\tau_m)} \right)
\end{equation}
where we adopt $f_p = -1$ for orbital expansion associated
with conservative evaporation and $f_p >0$ for the possibility
of orbital decay induced by
mass and angular momentum loss from the systems.

We incorporate the effect of planets' mass-loss by introducing a
torque in their equation of motion
\begin{eqnarray}
\frac{d}{dt}\textbf{V}_i =
 -\frac{G(M_*+m_i)}{{r_i}^2}\left(\frac{\textbf{r}_i}{r_i}\right)~~~~~~~~~~
 \nonumber\\
+\sum _{j\neq i}^N Gm_j \left[\frac{(\textbf{r}_j-\textbf{r}_i
)}{|\textbf{r}_j-\textbf{r}_i|^3}- \frac{\textbf{r}_j}{r_j^3}\right]
\nonumber\\
+\textbf{F}_\theta
+\textbf{F}_{\rm damp}+\textbf{F}_{\rm migI}, ~~~~~~~~
\label{eq:eqf}
\end{eqnarray}
where $\textbf{F}_\theta ={ {F}_\theta }{\hat {\bf \theta}}$ is being applied
at the radius $r_i$ in the azimuthal (${\bf \theta}$) direction with
\begin{equation}
F_\theta=\frac{\dot J}{r_i}=\frac{J}{r_i}
\left(\frac{\dot a_p}{2a_p} \right).
\end{equation}
In the above equation, $J=\sqrt{G M_* a_p}$ is the planets' specific
angular momentum at the guiding center of their epicycles. Note that
$F_\theta$ vanishes for $t > > \tau_m$.  Two
additional forces $\textbf{F}_{\rm damp}+\textbf{F}_{\rm migI}$
are included in the equation of motion to take into account the planets'
interaction with their natal disks during their infancy in section 3 (see
\S\ref{sec:formation}), in this section we only consider the effect of $F_\theta$.

For the numerical simulations to be presented below, Equation (\ref{eq:eqf})
is implemented into the Rebound Code \citep{RT15}. The orbits of all
planets are assumed to be coplanar around a $M_\ast= 1 M_\odot$ host
star.  At the onset of the calculation, we assume the planets' orbits
to be circular with randomly distributed mean anomaly and argument
of pericenter.  Each model is run for a maximum time scale of $10^7$ yrs
or until the planets' orbits cross each other.

\subsection{Orbit crossing in multiple planetary systems}
\label{sec:orbitcrossing}
For illustrative purpose, we consider an idealized system of
nine equal-mass ($M_p$) planets.  At the onset of the calculation,
they are equally separated in logarithmic interval. The
normalization is set, such that the semi-major axis of the
third innermost planet is designated to be at 1AU and that of
the $i+1^{th}$ planet is scaled to be at
\begin{equation}
a_{i+1} = a_i  + k R_H,  \ \ \ \ \ \ \ \ \ 
\end{equation}
where $R_H = (2 \mu / 3)^{1/3} (a_i + a_{i+1})/2$ is the Hill radius, $\mu=M_p/M_\ast$, and $k$ is a constant model parameter which represents the planetary separation in units of mutual Hill radii.  One of the planets in the system is designated to lose
a fraction $A/(1+A)$ of its initial mass due to the photo-evaporation
and to undergo either inward or outward migration over a fraction
$\sim -2 f_p A/(1+A)$ distance.

In these equal-mass systems, the innermost planet is the most likely planet
to endure photo-evaporation. For the first set of simulations, we
set $A=0.05$ and assume conservative photo-evaporation with $f_p=-1$ (i.e.
planets undergo outward migration as they loss mass).  We consider three
groups of simulations with $M_p=1$, 3, and 10 $M_\oplus$, respectively. For
each group, 12 models are computed with different values of $k$ (initial
separation) and mass-loss time scale ($\tau_m$).

\begin{figure*}[htp]
\begin{center}
  \epsscale{1.1}
  \plotone{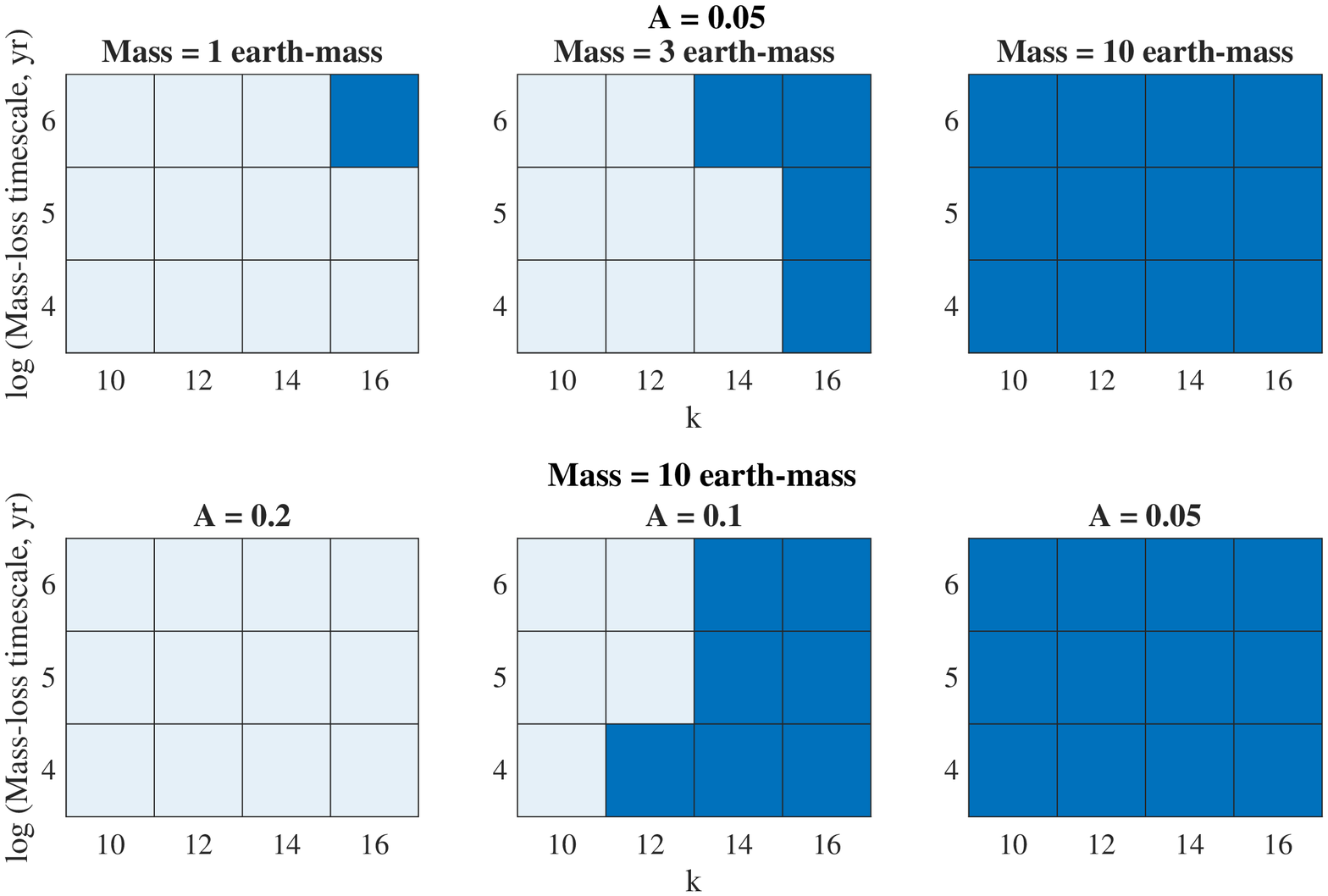}
 \caption{The stabilities of multiple planetary systems are shown with varied mass-loss timescales and fractions. k represents the relative Hill Radius between two adjacent planets initially. The y-axis represents the logarithm of mass-loss timescale.  The systems which are stable for at least $10^7$ yrs are labelled in dark blue, while the systems with orbital crossing before the end of simulations are labelled in light blue. Upper three panels show the stabilities of systems with different planetary masses, but with same mass-loss fraction (A=0.05). The cases in the upper three panels are corresponding to that listed in table \ref{allcase}. The lower three panels show the stabilities of systems with different mass-loss fraction, but with same planetary masses ($m=10 M_\oplus$). The cases with A=0.2 and A=0.1 are corresponding to that listed in table \ref{high}. The results shown in the upper and lower panels of the rightmost column are the same cases.}
 \label{stable}
 \end{center}
\end{figure*}

We list the initial parameters in Table \ref{allcase}.  Stable
models (planets avoided orbit crossing for at least $10^7$ yr)
are denoted in red color whereas models which led to orbit crossing
for any adjacent planets within $10^7$ yr are marked in black color. The stabilities of the systems in Table \ref{allcase} are also shown in the upper panels of Figure \ref{stable}.
The low-mass ($1 M_\oplus$) group have relatively small $R_H$
such that some compact systems (with modest $k$) may have
overlapping MMRs.  Only systems with large initial $k (\geq 16$)
and long mass-loss time scale $\tau_m \geq 10^6$ remain stable
for more than $10^7$ yrs.

\begin{figure}[htp]
\begin{center}
 \epsscale{1.2}
 \plotone{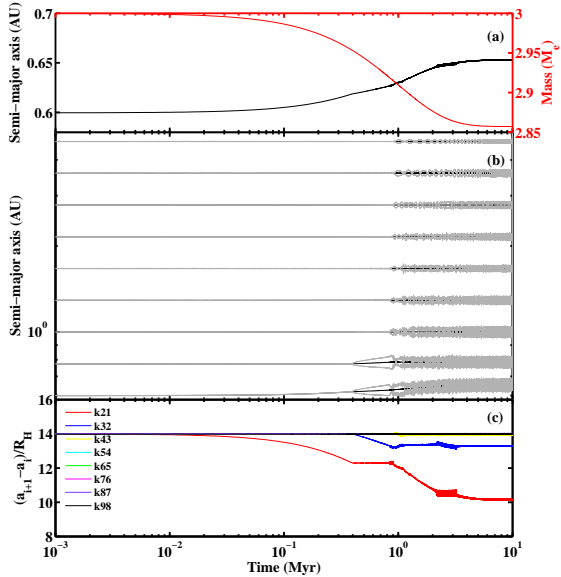}
 \caption{The evolution of a typical stable system. In this run, the masses of planets in the system are 3 $M_\oplus$, the initial separation between planets is 14 $R_{hill}$, and the timescale of mass-loss is $10^6$ yrs. The system is stable for more than 10 Myrs. The innermost planet loses 5\% of its total mass in this process. Panel (a) shows the evolution of semi-major axis and mass of the innermost planet, he red line shows the evolution of mass and the black line represents the evolution of semi-major axes. Panel (b) represents the semi-major axis evolution of nine planets in the system, the black lines mean the evolution of the semi-major axis and the grey lines display the evolution of the pericenter and apocenter. Panel (c) displays the evolution of the relative space $k_{ji}$ between two adjacent planets $P_i$ and $P_j$ (here, i represents the serial number of the planet from inner to outer, $1\leqslant i \leqslant 8$, $j=i+1$), different colors display different $k_{ji}$ as shown in the legend in panel (c).}
 \label{d6}
 \end{center}
\end{figure}

In comparison with the $M_p = 1 M_\oplus$ planets, the magnitude
of $R_H$ is larger for the $M_p= 3 M_\oplus$ planets.  Adjacent
planets with $k \geq 16$ are stable because they avoid most MMR's,
regardless the mass-loss time scale $\tau_m$.  In Figure
\ref{d6}, we show the evolution of a typical stable (for $>10^7$ yrs)
system (with $M_p=3 M_\oplus$, $A=0.05$ and $\tau_m=10^6$ yrs).  As a
consequence of innermost planet's outward migration, it is captured
into a 4:3 MMR with the next planet. For the high mass ($M_p =10
M_\oplus$), $R_H$ is sufficiently large, all systems with $k \geq 10$
are stable for more than $10^7$ yrs. In general, systems with larger
$M_p$, longer $\tau_m$, larger initial $k$ values are more stable.

\begin{table*}[htp]
\centering
 \caption{The initial parameters of the cases in section 2.2. A is the mass-loss fraction of the total planet (A=0.05 for the cases in this table), Mass represents the initial mass of planet in the system, $k$ is the initial planetary separation in units of mutual Hill radii, $\tau_m$ is the mass-loss timescale. The cases shown in red color are stable for at least $10^7$ yrs, while the cases in black represent the systems with orbital crossing happened before the end of simulations.}
\begin{tabular*}{16cm}{@{\extracolsep{\fill}}cc|ccc|ccc|ccc}
\tableline
\tableline
Case&A&Mass&$k$&$\tau_m$ &Mass&$k$&$\tau_m$ &Mass&$k$&$\tau_m$\\
&&$M_\oplus$&$R_{hill}$&yr&$M_\oplus$&$R_{hill}$&yr&$M_\oplus$&$R_{hill}$&yr\\
\tableline
1&0.05&1&10&$10^4$&3&10&$10^4$&\textcolor{red}{10}&\textcolor{red}{10}&\textcolor{red}{$10^4$}\\
2&0.05&1&10&$10^5$&3&10&$10^5$&\textcolor{red}{10}&\textcolor{red}{10}&\textcolor{red}{$10^5$}\\
3&0.05&1&10&$10^6$&3&10&$10^6$&\textcolor{red}{10}&\textcolor{red}{10}&\textcolor{red}{$10^6$}\\
4&0.05&1&12&$10^4$&3&12&$10^4$&\textcolor{red}{10}&\textcolor{red}{12}&\textcolor{red}{$10^4$}\\
5&0.05&1&12&$10^5$&3&12&$10^5$&\textcolor{red}{10}&\textcolor{red}{12}&\textcolor{red}{$10^5$}\\
6&0.05&1&12&$10^6$&3&12&$10^6$&\textcolor{red}{10}&\textcolor{red}{12}&\textcolor{red}{$10^6$}\\
7&0.05&1&14&$10^4$&3&14&$10^4$&\textcolor{red}{10}&\textcolor{red}{14}&\textcolor{red}{$10^4$}\\
8&0.05&1&14&$10^5$&3&14&$10^5$&\textcolor{red}{10}&\textcolor{red}{14}&\textcolor{red}{$10^5$}\\
9&0.05&1&14&$10^6$&\textcolor{red}{3}&\textcolor{red}{14}&\textcolor{red}{$10^6$}&\textcolor{red}{10}&\textcolor{red}{14}&\textcolor{red}{$10^6$}\\
10&0.05&1&16&$10^4$&\textcolor{red}{3}&\textcolor{red}{16}&\textcolor{red}{$10^4$}&\textcolor{red}{10}&\textcolor{red}{16}&\textcolor{red}{$10^4$}\\
11&0.05&1&16&$10^5$&\textcolor{red}{3}&\textcolor{red}{16}&\textcolor{red}{$10^5$}&\textcolor{red}{10}&\textcolor{red}{16}&\textcolor{red}{$10^5$}\\
12&0.05&\textcolor{red}{1}&\textcolor{red}{16}&\textcolor{red}{$10^6$}&\textcolor{red}{3}&\textcolor{red}{16}&\textcolor{red}{$10^6$}&\textcolor{red}{10}&\textcolor{red}{16}&\textcolor{red}{$10^6$}\\
\hline
\hline
\label{allcase}
\end{tabular*}
\end{table*}

\begin{table*}[htp]
\centering
 \caption{The initial parameters and results in the cases with high $A$ (A=0.1) in section 2.2. The definitions of parameters listed in the table are the same as that in table \ref{allcase}. The cases shown in red color are stable for at least $10^7$ yrs, while the cases in black represent the systems with orbital crossing happened before the end of simulations.}
\begin{tabular*}{10cm}{@{\extracolsep{\fill}}cc|ccc|ccc}
\tableline
Case&Mass&A&$k$&$\tau_m$ &A&$k$&$\tau_m$\\
&$M_\oplus$&&$R_{hill}$&yr&&$R_{hill}$&yr\\
\tableline
\tableline
1&10&0.1&10&$10^4$&0.2&10&$10^4$\\
2&10&0.1&10&$10^5$&0.2&10&$10^5$\\
3&10&0.1&10&$10^6$&0.2&10&$10^6$\\
4&10&\textcolor{red}{0.1}&\textcolor{red}{12}&\textcolor{red}{$10^4$}&0.2&12&$10^4$\\
5&10&0.1&12&$10^5$&0.2&12&$10^5$\\
6&10&0.1&12&$10^6$&0.2&12&$10^6$\\
7&10&\textcolor{red}{0.1}&\textcolor{red}{14}&\textcolor{red}{$10^4$}&0.2&14&$10^4$\\

8&10&\textcolor{red}{0.1}&\textcolor{red}{14}&\textcolor{red}{$10^5$}&0.2&14&$10^5$\\

9&10&\textcolor{red}{0.1}&\textcolor{red}{14}&\textcolor{red}{$10^6$}&0.2&14&$10^6$\\

10&10&\textcolor{red}{0.1}&\textcolor{red}{16}&\textcolor{red}{$10^4$}&0.2&16&$10^4$\\
11&10&\textcolor{red}{0.1}&\textcolor{red}{16}&\textcolor{red}{$10^5$}&0.2&16&$10^5$\\

12&10&\textcolor{red}{0.1}&\textcolor{red}{16}&\textcolor{red}{$10^6$}&0.2&16&$10^6$\\
\tableline
\tableline
\label{high}
\end{tabular*}
\end{table*}

These results are generalized to cases with non-conservative mass-loss.
We consider the possibility that it may lead to inward migration with
$f_p=1$.  Under this assumption, planetary migration would be divergent
and the systems would remain stable if the mass-losing planet has the
smallest semi-major axis.  But,  if the mass-losing planet is one of
the more distant member of a closely-spaced multiple planetary system,
non-conservative mass-loss would lead to convergent migration and
the onset of dynamical instability.  We have also examined the
possibility that the mass-losing planet is initially located in the
middle of the pack.  In this case, either inward or outward migration
can destabilize the system in accordance with the results presented in
Table \ref{allcase}.

The gaseous envelope contributes $\sim 10 \%$ of the total mass in
Uranus and Neptune.  We consider a second series of models with larger
fractional mass-loss ($A=0.1$ and 0.2) from the planets with
$M_p=10 M_\oplus$, $f_p=-1$ (outward migration), a range of values
for $k$ and $\tau_m$ (Table \ref{high}).  These models indicate that
greater fractional mass-loss (ie larger $A$) leads to more extended
migration and forces the planetary system to evolve into more compact
configuration and become unstable, as shown in the bottom panels of Figure \ref{stable}. Modest mass-loss ($A=0.2$) can
destabilize the systems even in the limit that they started with a wide
initial separation ($k =16$) regardless the masses of the planets and
the mass-loss time scale.

\section{Planetary systems with MMRs}
\label{sec:planetmmrs}

The results in the previous section clearly show
the influence of mass-loss on the stability of 
multiple planetary systems.  However, the equal-mass systems we have adopted as initial conditions are
highly idealized.  Emerging multiple planets around the same host stars are likely to adjust their initial separation due to their interaction with
their natal disk and each other. Their differential
migration may lead them to be captured into each other's MMRs.

In this section, we consider the impact of migration induced by
either impulsive giant impacts or gradual photo-evaporation on
some compact multiple planetary systems with MMRs.  For initial
conditions, we first construct a generic formation model for these
systems in \S\ref{sec:observation}.  Applying modest amount of
uniform compression or expansion, we carry out numerical
experiments to show that although equilibrium configuration
may emerge after the disk depletion,  their islands of stability
are narrowly confined.  Due to this effect, minor impulses
and/or protracted loss of a modest amount of mass can
indeed lead to dynamical instabilities for these systems.

\subsection{Multiple planetary systems with MMRs}
\label{sec:observation}
Resonant systems are common, $\sim 20\%$ and $\sim 10\%$ of Kepler's
multiple-planet systems are near their 2:1 and 3:2 MMRs,
respectively \citep{Wang12, lee13, Liu15,  Wang14, Wang17, Wang21}.
Chain resonances with period ratio in 4:2:1 MMRs or 6:3:2
MMRs have been found in some triple systems. Two sets of
8:6:4:3 and 6:4:2:1 resonant chains are found among four super-Earths around Kepler-223 \citep{De17, Mills16} and Kepler-79
\citep{Wang12}, respectively. They have masses $\sim 3.5-7~M_\oplus$.
Five planets, separated by their mutual 3:2 MMR's, are found
with radii ($1.6-3.2~R_\oplus$) in K2-138 system on either side of size gap
\citep{Christ18}. Six out of seven planets in the TRAPPIST-1
system are in near chain resonances \citep{Gi17}. Five out of the six planets in the TESS mission found system TOI-178 are in near chain resonances as shown in Figure \ref{multi} \citep{Leleu2021}.

Through exhaustive numerical integration, the present-day
masses and orbital elements of these systems have been determined, under the constraint that they are stable for at least $10^7$ yrs
\citep{Tam17}. These calculations do not, however, take into account
the possibility of any mass-loss from one or more of their members, even
though some of these planets have observed radii smaller than the
lower bound of the gap in the super-Earths' radius distribution
(presumably they do not have any significant atmosphere, possibly as a result of previous photo-evaporative mass-loss).
Due to the vast range of potential initial orbital configurations
and mass-loss efficiency, it would not be practical to separately
examine the effects of mass-loss on all the individual observed
systems. Instead, we adopt some generic formation scenarios for MMR
systems and utilize the results generated from these models as the
initial conditions for the systems' subsequent impulsive or prolonged
orbital evolution.

Planets in observed chain MMRs or near MMRs systems have masses
several times that of the Earth.  Their orbital semi-major axes
are typically an order of magnitude larger than the radii of their
host stars.  {\it In-situ} formation
of these super-Earths requires disk surface density in planetesimals
to be many orders of magnitude larger than that extrapolated from the
minimum mass nebula scenario \citep{idalin04, Wang12, Chiang13, Tam17}.
Although high-eccentricity migration due to the planet-planet scattering
\citep{RF96, LI97, BN12}, Kozai resonance \citep{WM03, And16},
and the secular chaos \citep{Nag08, WL11, Ham17} can lead to the
relocation of close-in Jupiter-mass gas giants, the configurations
of MMRs in multiple super-Earth systems are difficult to accomplish
\citep{Giacalone17, Xue17} in these circumstances.

\subsection{Formation of multiple planets with MMRs}
\label{sec:formation}
The most plausible scenario for forming super Earth planetary system
with MMRs is orbital migration due to their tidal interaction with
their natal disks \citep{GT80, lin96, Bryden00, MS01, miga16, iz17, Liu17a,
Liu17b, ramos17}.  These planets' masses are too small to clear
deep gaps in the gas surface density distribution of their natal disks.
Nevertheless, they exert a tidal torque on the disk which leads to
planets' type I migration \citep{Ward97, Tan02}.  The direction
and pace of embedded planets' migration is determined by the total torque
\begin{eqnarray}
\Gamma_{\rm tot} (r_p)=f_{tot}\Gamma_0, \ \ \ \ \ \ \ \ \ {\rm where} 
\nonumber\\
\Gamma_0=(\mu/h)^2 \Sigma_p r_p^4 \Omega_p^2/\gamma,~~~~~~~~~
\end{eqnarray}
where $\gamma=1.4$ is the adiabatic index, $h$ is the aspect ratio,
and $\Omega_p$ is the Keplerian frequency at the location of the planet. We adopt the empirical minimum mass solar nebular model (MMSN;
Hayashi 1981) for $\Sigma_p=\Sigma_g (r_p)$ with
\begin{equation}
\Sigma_{\rm g} (r)
={1700 f_{\rm g} \over (r/1AU)} \rm exp\left(\frac{-t}{\tau_{\rm dep}}
\right)\rm g \ cm^{-2},
\label{gasdens}
\end{equation}
where $f_{\rm g}$ is the enhancement factor. Herein, we consider the gas
depletion time scale $\tau_{\rm dep}$ in the range of [0.1, 3] Myrs
\citep{hai01}.

The total-torque coefficient $f_{\rm tot} = f_{CR} + f_{LB}$ is the
sum of corotation ($f_{CR}$) and Lindblad ($f_{LB}$) torque on the
planets \citep{Paa10, Paa11}. The magnitude and sign of $f_{CR}$ and
$f_{LB}$ are functions of the evolving distribution of $s=\partial
{\rm ln} \Sigma_g / \partial {\rm ln} r$, $\beta = \partial {\rm ln} T
/ \partial {\rm ln} r$, viscosity $\nu$ and thermal diffusivity $\xi$.
For the mass range of super-Earths considered here, the corotation
torque is unsaturated and maximally effective.  It can lead to outward
type I migration in the viscously heated inner disk region and inward
migration in the irradiated outer disk region \citep{KL12}. Consequently,
the migrating super-Earths have a tendency to converge toward the
transition radius $r_t$ which separates these two domains.

In order to take into account these possibilities, we introduce a
prescription to characterize this transition in $f_{CR}$ across
$r_t$ such that
\begin{equation}
f_{CR}={\rm coef} \times
\left( 1-\frac{2(r/r_t)^2}{1+(r/r_t)^2} \right),
\end{equation}
where we consider a range for the torque coefficient ${\rm coef}$
(2, 5, or 10).  While $r_t \sim$ a few AU in a MMSN, it can reduce
to a fraction of 1 AU during the disk depletion phase
\citep{GaraudLin2007, Wang21}.  For our simulations, we choose two sets of
$r_t = 0.5$ and $3$ AU to represent the range of possibilities.

In most regions of the disk, the Lindblad torque is negative ($f_{LB}
< 0$).  But, near their inner edge $r_m$, disks are sharply truncated by
the magnetosphere of their host stars and a steep positive surface
density gradient leads to super Keplerian azimuthal velocity.  In
this region the Lindblad torque becomes positive ($f_{LB} > 0$) such
that inward migration of super-Earths may be stalled. In order to take
into account this possibility, we introduce another prescription for
\begin{equation}
f_{LB}=1-\frac{2(r/r_m)^4}{1+(r/r_m)^4}.
\end{equation}

Tidal torque due to disk-planet interaction leads to type I migration on a
time scale
\begin{equation}
\tau_a = {a \over {\dot a}} = {M_p \sqrt{ G M_\ast r} \over 2
\Gamma_{\rm tot} } = {\tau_0 (r_p) \over f_{\rm tot}},
\end{equation}
where $\tau_0 = M_\ast (h^2 \gamma / 2 \mu \Omega_p) /(\Sigma_p r_p^2$)
\citep{Tan02,KN12}. The associated time scale for eccentricity damping is
$\tau_e = e/{\dot e} = h^2 \tau_a$.  These contributions are incorporated
into the Equation of motion (Eq \ref{eq:eqf}) through
\begin{equation}
\textbf{F}_{\rm migI} = { \Gamma_{\rm tot} {\hat {\bf \theta}} \over M_p r }
\ \ \  {\rm and}
\end{equation}
\begin{equation}
\textbf{F}_{\rm damp} = - { ({\bf v \cdot r}) {\bf r}  \over r \tau_e}.
\end{equation}
Aside from $\textbf{F}_{\rm migI}$ and $\textbf{F}_{\rm damp}$, we do not take into account
the self gravity of the disk gas nor its effect
on the procession of the planets' orbits. While the
former contribution is negligible, the latter effect
may lead to secular resonances among well separated
planets in sparsely populated systems \citep{Zheng2017}.

\subsection{Emergence of stable multiple planetary systems with MMRs}
\label{sec:origin}
In order to obtain some reasonable realistic initial conditions, we
adopt a generic formation scenario for multiple planetary systems with
MMRs based on the assumption that their kinematic orders were established
through differential and convergent type I migration \citep{Liu17a, Liu17b}.
Similar to the previous section, we adopt systems of nine equal mass
($M_p=10^{-5} M_\odot$) planets designated to be P1 to P9 in order to
increasing $a$ around a $M_\ast=1 M_\odot$ \citep{Zhou07}. At the
onset of our calculation, we place these planets with equal fractional
spacing $k R_H$ around the third (Earth-like) planet at 1AU.

In order to examine how planets' asymptotic semi-major axis and
eccentricity distribution may depend on the evolution of disk properties
\citep{Wang17}, we consider three groups of models (G1, G2, G3) with
several different values of $f_g (=0.1, 0.2, 0.5, 1)$ and a range
of $\tau_{\rm dep} (= 0.1, 1/3, 1, 3$ Myr).  For the planets, we
adopt the same value for ${\rm coef} (=10)$ and $r_m (=0.05$ AU)
for two initial sets of $k (= 8$, 16) and $r_t (=0.5, 3$ AU) (Table
\ref{run1}).  These model parameters lead to three representative
groups of multi-planet systems with diverse MMRs configurations:

\begin{table*}
\small
\centering \caption{Parameters used in the simulations, Group 1, Group 2, and Group 3. $k_0=(a_{i+1}-a_i)/R_H$ represents the relative separation between two adjacent planets initially. $r_{\rm t}$ is the transition radius, $\tau_{\rm dep}$ is the depletion timescale of the gas disk, $coef$ is the coefficient in the corotation torque, h=H/r represents the disk's aspect ratio, and $f_g$ is the enhancement factor of the gas density.}
\begin{tabular*}{16cm}{@{\extracolsep{\fill}}ccccccc}
\tableline
\tableline
 & $k_0$&$\tau_{\rm dep}$& $r_{\rm t}$&$coef$&$h=H/r$&$f_{\rm g}$\\
&$R_{\rm hill}$& Myr & AU& & \\
\tableline
G1&8&1&0.5&10&$0.05r^{0.25}$&0.1\\
G2&16&1&3.0&10&0.05&0.1\\
G3-1&16&1/3&3.0&10&0.05&1\\
G3-2&16&1/3&3.0&10&0.05&0.5\\
G3-3&16&1/3&3.0&10&0.05&0.2\\
G3-4&16&0.1&3.0&10&0.05&0.2\\
G3-5&16&1&3.0&10&0.05&0.2\\
G3-6&16&3&3.0&10&0.05&0.2\\
\tableline
\tableline
\label{run1}
\end{tabular*}
\end{table*}

\begin{table*}[htp]
 \caption{The orbital parameters of the planets at the end of the simulation in the case in Group 2, which also represent the  initial parameters of planets in the cases in section 3.2. The planets are named with their serial number from the inner to outer with $a$, $e$, $i$, $\Omega$, $\omega$, $f$, and Mass representing the semi-major axis, eccentricity, inclination, longitude of the ascending node, argument of periastron, true anomaly and mass of planets, respectively.}
\begin{tabular*}{16cm}{@{\extracolsep{\fill}}ccclllll}
\tableline
Name&$a$&$e$&$i$&$\Omega$& $\omega$&$f$&Mass\\
&AU& &rad&rad& rad & rad&$M_\odot$\\
\tableline
1&0.753710&0.058854&0&0&0.387958&-1.367171&1e-5\\
2&0.987895&0.102676&0&0&-2.749728&-0.024010&1e-5\\
3&1.293961&0.077092&0&0&0.362406&-0.952595&1e-5\\
4&1.696034&0.064311&0&0&-2.754476&-1.830721&1e-5\\
5&2.222124&0.044478&0&0&0.591902&-2.323546&1e-5\\
6&2.912175&0.039683&0&0&-2.227325&-2.776581&1e-5\\
7&3.820460&0.024706&0&0&-0.013042&-0.058686&1e-5\\
8&4.623438&0.024523&0&0&2.482278&0.533084&1e-5\\
9&5.608244&0.011006&0&0&-0.760252&2.023750&1e-5\\

\tableline
\label{run2}
\end{tabular*}
\end{table*}

\begin{figure}
\begin{center}
  \epsscale{1.3}
  \plotone{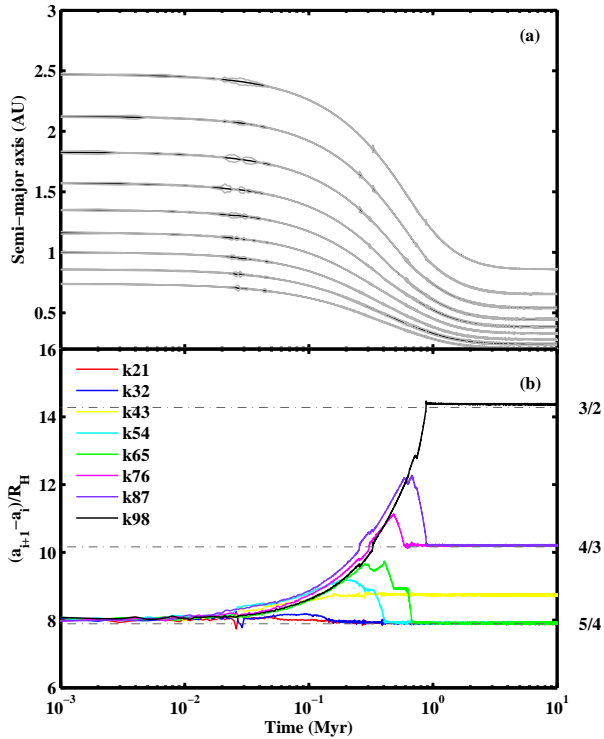}
 \caption{The results of G1. In this Group, all planets undergo inward type I migration. The transition radius is 0.5 AU. The planets are labelled in 1 to 9 from the innermost to the outermost. Panel (a) shows the evolution of semi-major axes of nine planets with black lines, the evolution of the pericenter and apocenter of nine planets are shown in grey lines. Panel (b) displays the evolution of the relative separation between planets. k$ji$ means the the relative separation between two adjacent planets P$i$ and P$j$ (here, i represents the serial number of the planet from inner to outer, $1\leqslant i \leqslant 8$, $j=i+1$), different colors display different $k_{ji}$ as shown in the legend in panel (b).
 \label{G1}}
 \end{center}
\end{figure}

\textbf{Group 1}:
All the planets start with $a_i > r_t$ and initial separation of
8 $R_H$.  In the absent of gas, their orbits would be stable for
at least $10^7$ yrs \citep{Zhou07}.  In a low-mass ($f_g =0.1$) thin
($h=0.05(r/1\rm AU)^{0.25}$) disk with $r_t (=0.5$ AU), all planets
undergo type I inward migration (panel (a), Figure \ref{G1}). The
light grey lines represent the planets' apo and peri center distances.
The innermost planet first stalls at $r_t$. Others follow and they force
it to a smaller radius. Convergent migration leads to asymptotic
separation in the range of 8-14.2 $R_H$. Several planets capture
each other into a chain of discrete (3:2, 4:3, and 5:4) MMRs (panel b)
with integer period ratios and small eccentricity ($e \leq 0.035$).
Planets P5, P6, and P7 form a 5:4:3 resonant chain and P7, P8, and P9
form a 4:3:2 MMRs. Therefore, according to the estimation of crossing time \citep{Zhou07}, such configuration formed under inward migration with relative separation larger than 8 $R_H$ will be stable for more than $10^7$ yrs.

\begin{figure*}
\begin{center}
  \epsscale{1.0}
  \plotone{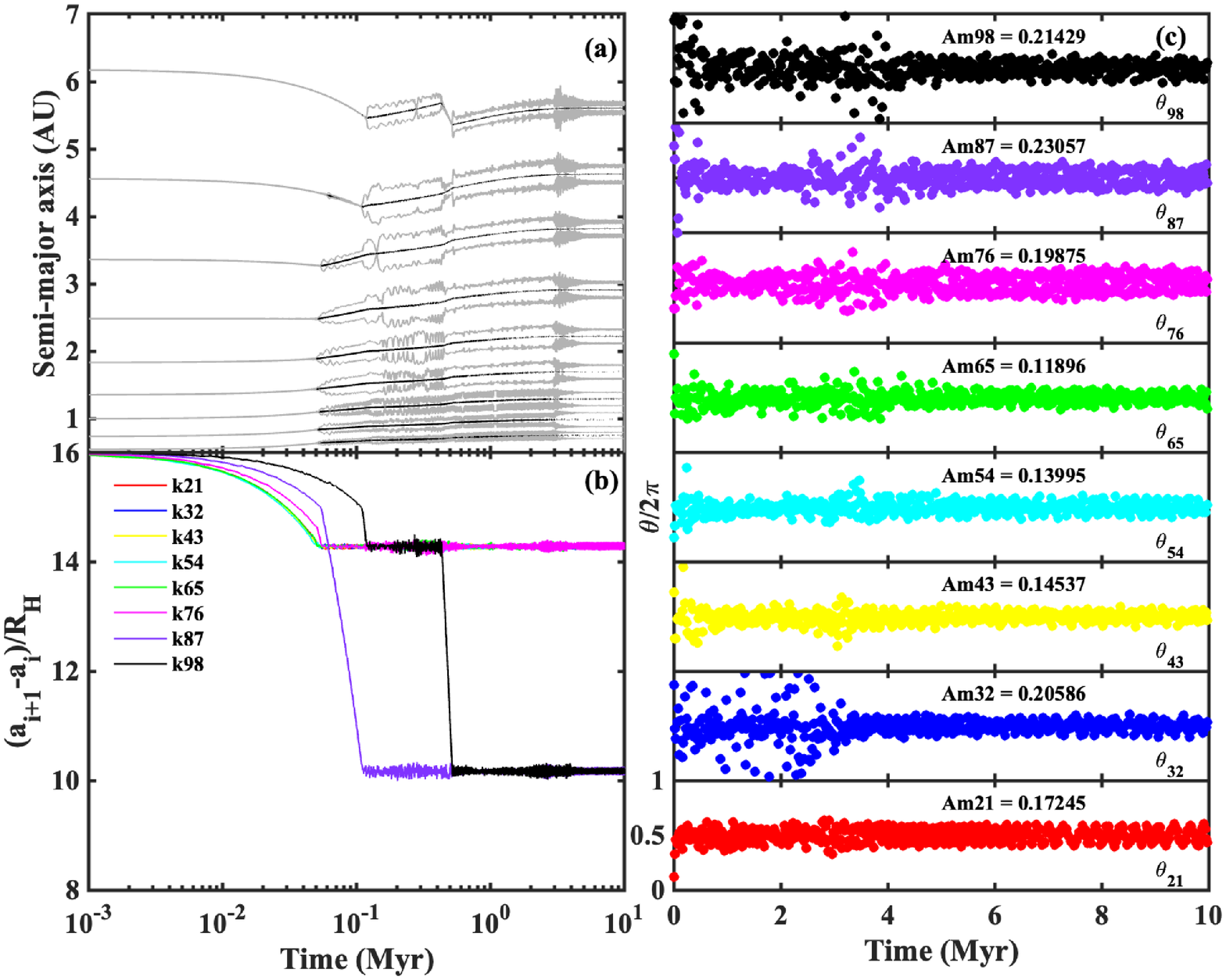}
 \caption{The results of G2. In this Group, the transition radius locates at 3 AU. Thus planets 7, 8 and 9 undergo inward migration, while the other planets migrate outward. Planet pairs can be trapped into 3:2 ($k$=14.2) and 4:3 MMRs ($k$=10.1). Panel (a) shows the evolution of semi-major axis, pericenter and apocenter of nine planets. The black lines display the evolution of semi-major axis, and the grey lines show the evolution of pericenter and apocenter of nine planets. Panel (b) shows the evolution of relative space between adjacent planet $k_{ji}$, the meanings of $k_{ji}$ are the same as in Figure \ref{G1}. Panel (c) shows the evolution of resonant angles $\theta=(p+q)\lambda'-p\lambda-q\varpi'$, where the unprimed and primed parameters refer to the inner and outer planets respectively, two planets are in or near (p+q):p resonance, $\lambda$ and $\varpi$ are the mean longitude and longitude of pericentre. Am$ji$ represents the resonant angle amplitude of the planet pair $P_i$ and $P_j$ (i represents the serial number of the planet from inner to outer, $1\leqslant i \leqslant 8$, $j=i+1$).
 \label{G2}}
 \end{center}
\end{figure*}

\textbf{Group 2}: Widely separated ($16 R_H$) planets are installed
on either sides of $r_t (=3$ AU) in a low-mass ($f_g =0.1$) disk.
In a gas-free environment, their orbit crossing time is considerably
$>10^7$ yrs \citep{Zhou07}.  In their natal disks, planets 7, 8, and
9 undergo inward type I migration, while other planets migrate outward.
Panels (a) and (b) in Figure \ref{G2} show the evolution of $a_i$ and
the separation between adjacent planets. After $10^5$ yrs, the inner
six planet pairs are captured into 3:2 MMRs, and planet 7 and 8 are
in 4:3 MMR. P9 is also captured into a 3:2 MMR at about $2\times
10^5$ yrs with P8 firstly, and into a 4:3 MMR at $\sim 5\times 10^5$ yrs.
Planets P8, P7, and P6 are also trapped in a 4:3:2 chain MMRs.  All the resonant angles librate with
small amplitude after the planets are trapped into MMRs (Panel c) with
the exception of that between P7 and P8 ($\sim 83^\circ$). Their
eccentricities are $\sim 0.1$ (see Table \ref{run2}). Different from the results in Group 1, the relative separation between two adjacent decreases with their evolution. The multiple planetary system may become unstable after the inward and outward migration. 

\begin{figure}
\begin{center}
  \epsscale{1.3}
  \plotone{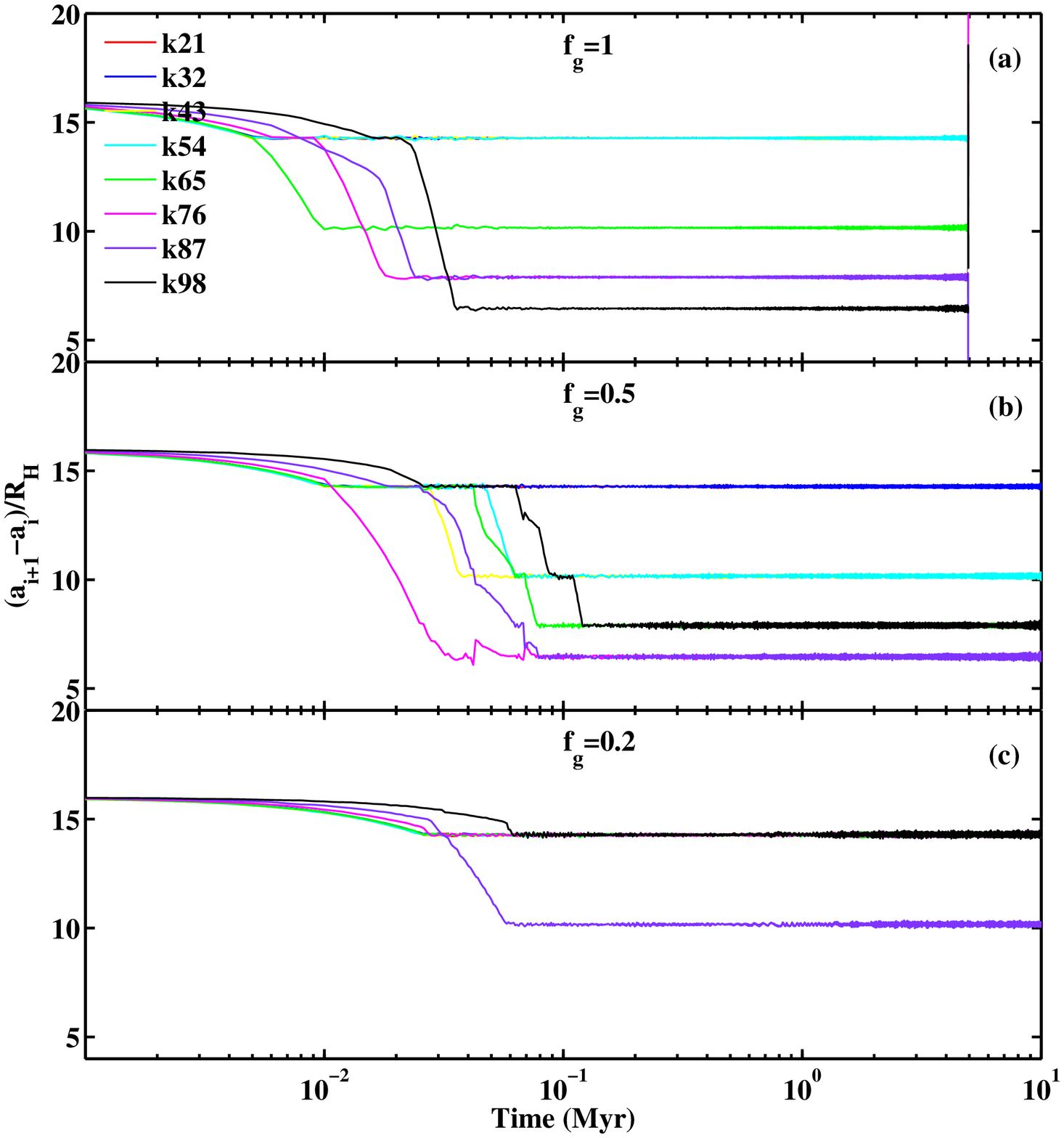}
 \caption{The evolution of relative separation between adjacent planets in different cases with different gas density. Panel (a), (b), and (c) are corresponding to the cases with $f_g$=1, 0.5, and 0.2, respectively. $f_g$ is the enhancement factor of the gas density as shown in equation (\ref{gasdens}). $f_g=1$ refers to the gas density profile of MMSN model. The colors of different lines represent the relative separation between different planet pairs as the legend shown in panel (a).
 \label{dens}}
 \end{center}
\end{figure}

\begin{figure}
\begin{center}
  \epsscale{1.2}
  \plotone{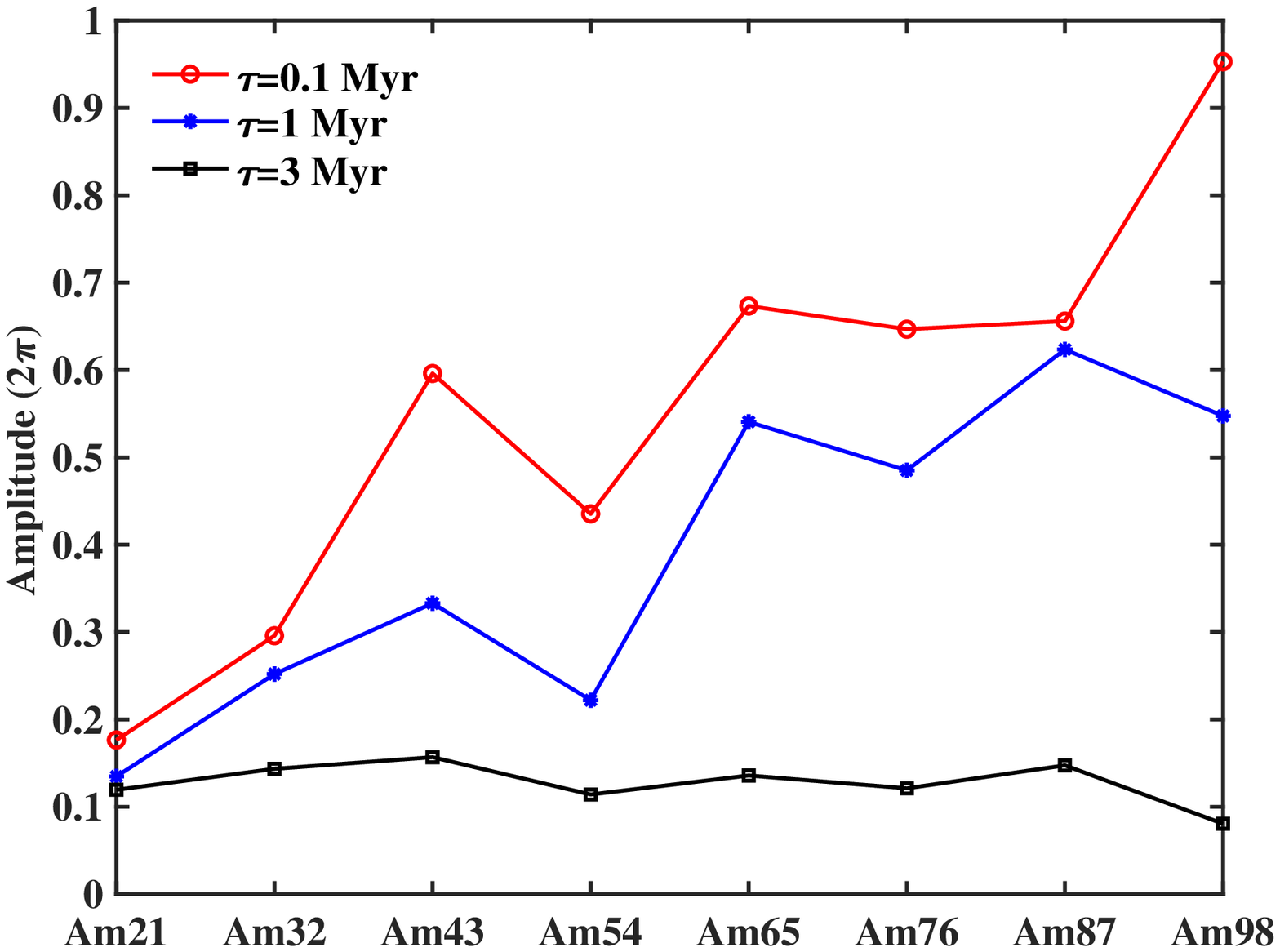}
 \caption{The amplitude of resonant angles in the system with different gas disk depletion timescales. Am$ji$ ($1\leqslant i \leqslant 8$, $j=i+1$) has the same definitions as in Figure \ref{G2}.  Red circles, blue asterisks and black squares represent depletion timescales of $10^5$, $10^6$ and $3\times 10^6$ yrs respectively.
 \label{am}}
 \end{center}
\end{figure}

\textbf{Group 3}: The initial conditions are similar to that in Group 2,
but with different $f_g (=0.2-1)$ and depletion timescale (0.1-3 Myr)
of the gas disk. In a minimum mass nebula (Model G3-1, panel (a) of
Figure \ref{dens}), the inner three planet pairs enter into
3:2 MMRs, P5 and P6 into 4:3 MMR, two pairs (P6 and P7, P7 and P8) into 5:4 MMRs.
P8 and P9 into 6:5 MMR in less than 0.1 Myr. This system becomes unstable
after $\sim 5$ Myrs. In a less-massive disk (Model G3-2 with
$f=0.5$ in panel b), the planets are captured into similar MMR's
in $\sim 0.1$ Myrs to that in G3-1. In the least-mass disk (Model G3-3 with $f_g
= 0.2$ in panel c), planet pairs enter into 3:2 and 4:3 MMRs with
wider spacing between them. Thus, with a less dense gas disk, the configuration formed through orbital migration can be stable for longer time with larger relative separations \citep{Zhou07}. In comparison with Model G3-3, we
consider $\tau_{\rm dep} = 0.1, 1, 3$ Myrs in models G3-4, G3-5,
and G3-6.  With a longer depletion timescale, the asymptotic
configuration is more stable with smaller
libration amplitude of their resonant angles (Figure \ref{am}).

The above results show the robust emergence of multiple planetary systems
with MMRs with different disk properties, and reveal the relationship between the stability of configuration formed through orbital migration and the disk properties.  Here, we utilize these
results as initial conditions to study the consequence of mass-loss
and its associated orbital evolution.

\subsection{Island of stability near the MMRs}
\label{sec:island}
In this section, we adopt an idealized prescription to
demonstrate the limited extent of stable region in the proximity of a set
of representative MMRs from the results of Model G2
(Figure \ref{G2}).  First, we uniformly redistribute
the planets' azimuthal position without any change to
their semi-major axes.  Since the eccentricities of
all the planets are very small, these changes correspond
to the randomization of their resonant angles.  For some
pairs, this modification breaks their MMR since the
modified resonant angles are outside their original range
of libration.  Nevertheless, these systems remain stable
without orbit crossing over a time scale of $10$ Myr.

In order to test the extent of these islands of stability, we
introduce an artificial, impulsive, uniform, fractional, expansion
or compression for all the planets in these systems.  For
computational simplicity, we multiply the separation between
adjacent pairs with a constant factor of $f_k <1$ or $f_k>1$
respectively. We fix the semi-major axis of the middle
planet (P5) and study the implications of artificial incremental changes
with $f_k$ in the range between 0.8 and 1.25. This approach,
though idealized, provides suggestive indication on the limited
extent of islands of global stability in closely packed multiple
systems.

\begin{figure}
\begin{center}
  \epsscale{1.2}
  \plotone{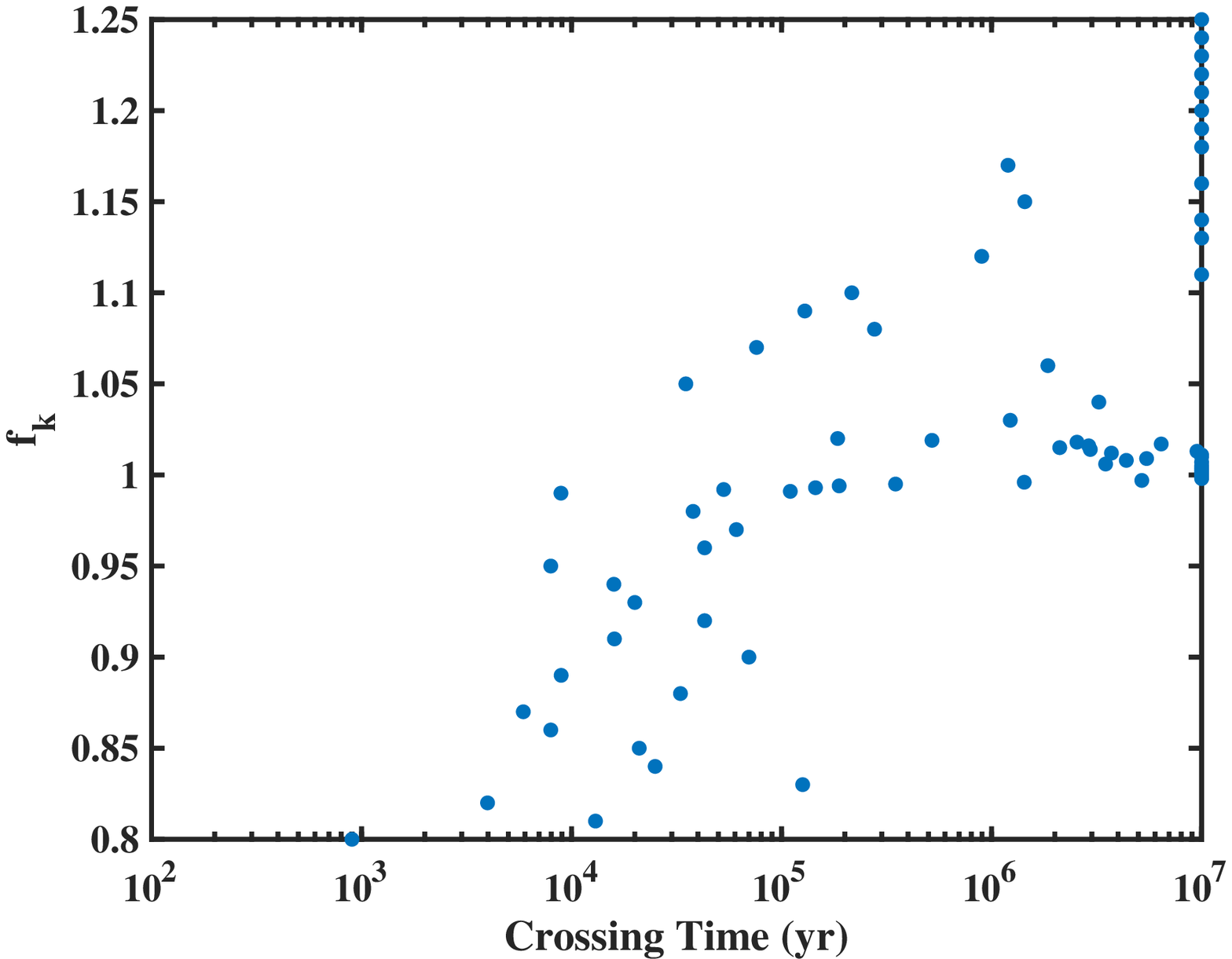}
 \caption{The distribution of crossing time vs the separation change of planet pairs. The original system which is obtained from G2 can be stable for more than $10^7$ yrs as shown in Figure \ref{G2} . In order to find out the stable region for such multiple planetary system, we introduce an artificial factor $f_k$ acting on the original stable system to change the separation between adjacent planets. We fix the semi-major axis of the middle planet $P_5$ and change the separation between adjacent by $f_k$, $0.8 \leq f_k \leq 1.25$. $f_k<1$ refers to a global contraction comparing to the original system configuration and $f_k>1$ represents an expansion to the original stable system.
 \label{ct}}
 \end{center}
\end{figure}

The evolution of these systems is examined based on the numerical solution
of Equation (\ref{eq:eqf}) with ${F}_\theta$, $\textbf{F}_{\rm damp}$, and
$\textbf{F}_{\rm migI}$ are set to be zero.  The time scale for the first set
of planet to cross each other's orbit is plotted in Figure \ref{ct}.
In general, a relatively small global contraction (with $f_k$ slightly less
than unity) can reduce the crossing time scale to less than 1 Myr.  But, the
stability of this system is preserved (on time scale longer than 10 Myrs)
with either a very limited changes ($1\leq f_k\leq 1.01$) or a large
($f_k \geq 1.2$) expansion factor, although some intermediate
($1.01 < f_k < 1.1$) expansion factor can lead to relative short
($<2$ Myrs) orbit crossing time.

\begin{figure*}[htp]
\begin{center}
  \epsscale{1.1}
  \plotone{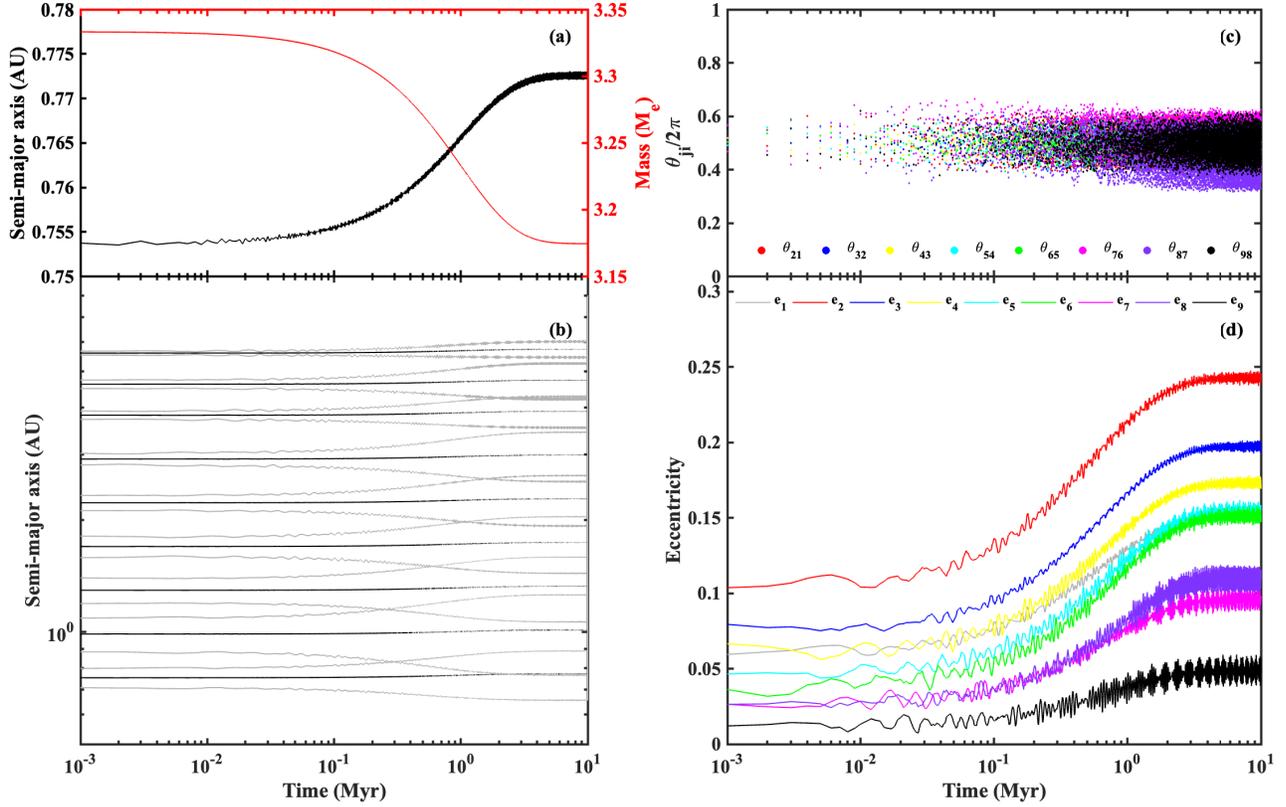}
 \caption{The evolution of the system in Case 1, A=0.05 and $\tau_m=10^6$ yrs.  Panel (a) shows the evolution of the innermost planet. The black line means the evolution of the semi-major axis and the red one represents the evolution of mass. In this case, the innermost planet will lose about 0.14 Earth-mass which is about 5\% of the initial mass. Panel (b) shows the evolution of semi-major axis, percenter and apocenter. The black lines mean the evolution of the semi-major axis and the grey lines display the evolution of the pericenter and apocenter. At the end of the simulation, the eccentricities of planets will be excited. But the system is still stable due to the relative angles between them. Panel (c) shows the evolution of the resonant angles $\theta_{ji}$ of two adjacent planet $P_i$ and $P_j$, $1\leqslant i \leqslant 8$, $j=i+1$. $\theta_{ji}=(p+q)\lambda'-p\lambda-q\varpi'$, where the unprimed and primed parameters refer to the inner and outer planets respectively, two planets are in or near (p+q):p resonance, $\lambda$ and $\varpi$ are the mean longitude and longitude of pericentre. Panel (d) shows the evolution of eccentricities of planets in the system $e_{pi}$, $pi$ represents the serial number of the planet from inner to outer, $1\leqslant pi \leqslant 9$, the eccentricities of different planets are labeled in different colors, as shown in the legend in panel (d). The evolution with A=0.1 and $\tau_m=10^6$ yrs in Case 2 is similar to that in Case 1.}
 \label{case1}
 \end{center}
\end{figure*}

\begin{figure*}[htp]
\begin{center}
  \epsscale{1.1}
  \plotone{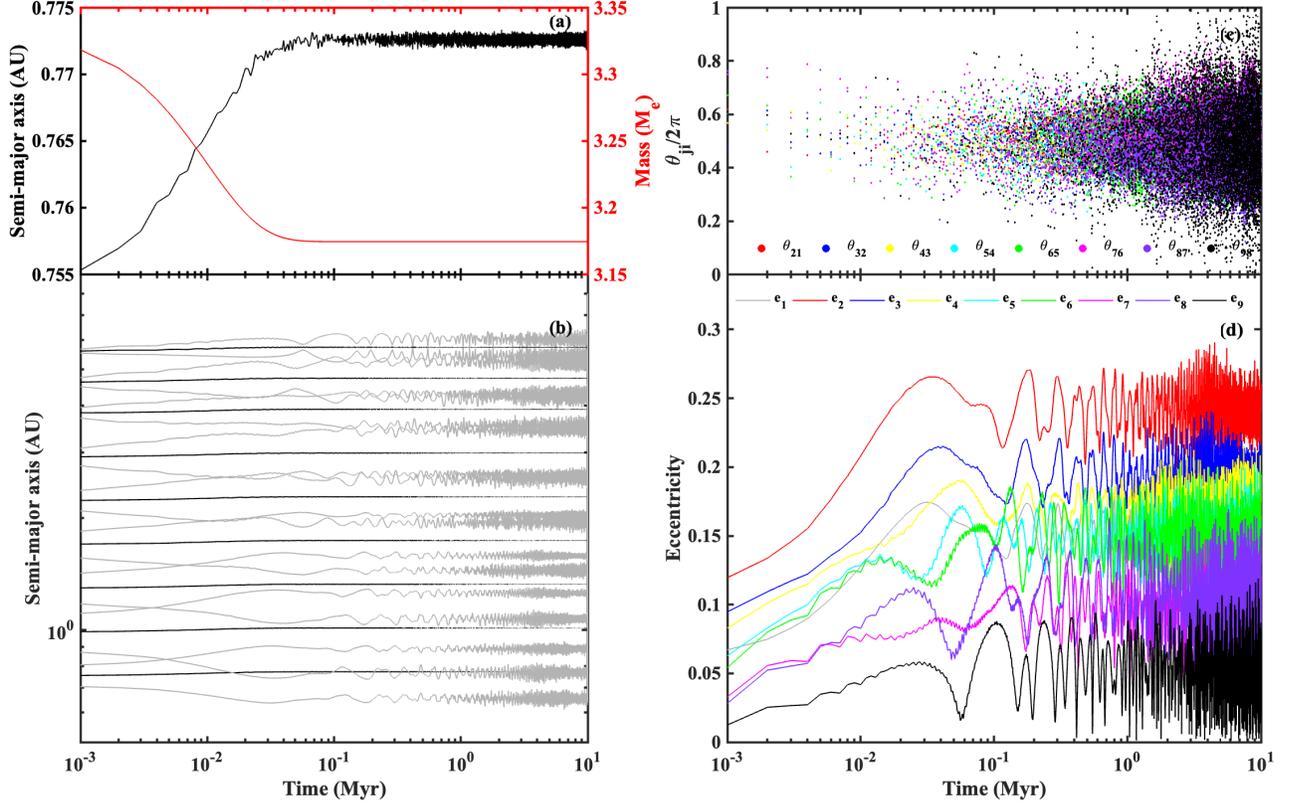}
 \caption{The evolution of the system in Case 3, A=0.05 and $\tau_m=10^4$ yrs.  Panel (a) shows the evolution of the innermost planet. In this case, the innermost planet will lose about 0.14 Earth-mass which is about 5\% of the initial mass. Panel (b) shows the evolution of a, a(1-e), and a(1+e) of all the planets. At the end of the simulation, the results are similar to that in Case 3. Panel (c) shows the evolution of the resonant angles. The resonant angles tend to be librate in a large range than in Case 1. Panel (d) shows the evolution of eccentricities of planets in the system. The meanings of different colors are the same as in Figure \ref{case1}}.
 \label{case3}
 \end{center}
\end{figure*}

\begin{figure*}[htp]
\begin{center}
  \epsscale{1.1}
  \plotone{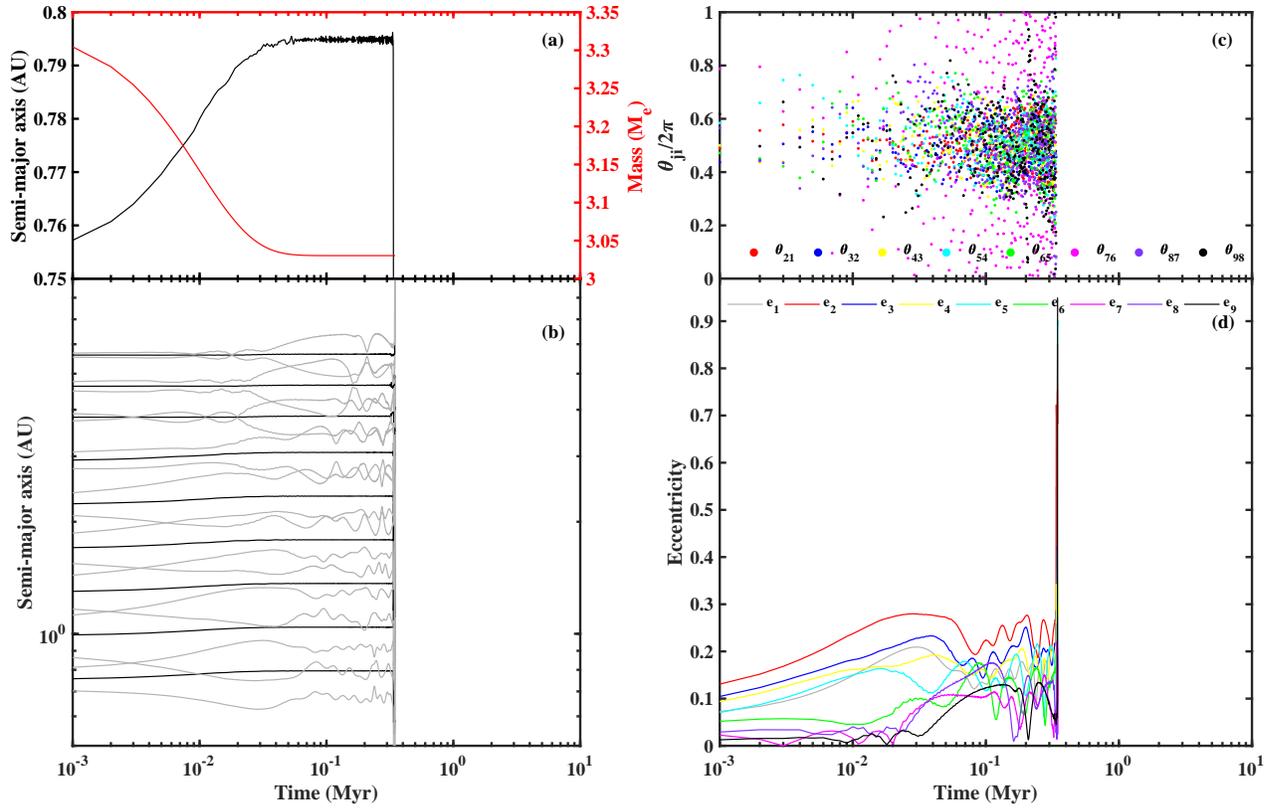}
 \caption{The evolution of the system in Case 6, A=0.1 and $\tau_m=10^4$ yrs. The system become unstable at about 0.35 Myrs. Panel (a) shows the evolution of the innermost planet. In this case, the innermost planet will lose about 0.3 Earth-mass which is about 10\% of the initial mass. Panel (b) shows the evolution of a, a(1-e), and a(1+e) of all the planets. Panel (c) shows the evolution of the resonant angles. The resonant angles tend to be librate in a large range from 0 to 2$\pi$. Panel (d) shows the evolution of eccentricities of planets in the system. The meanings of different colors are the same as in Figure \ref{case1}}.
 \label{case6}
 \end{center}
\end{figure*}

\begin{figure*}[htp]
\begin{center}
  \epsscale{1.1}
  \plotone{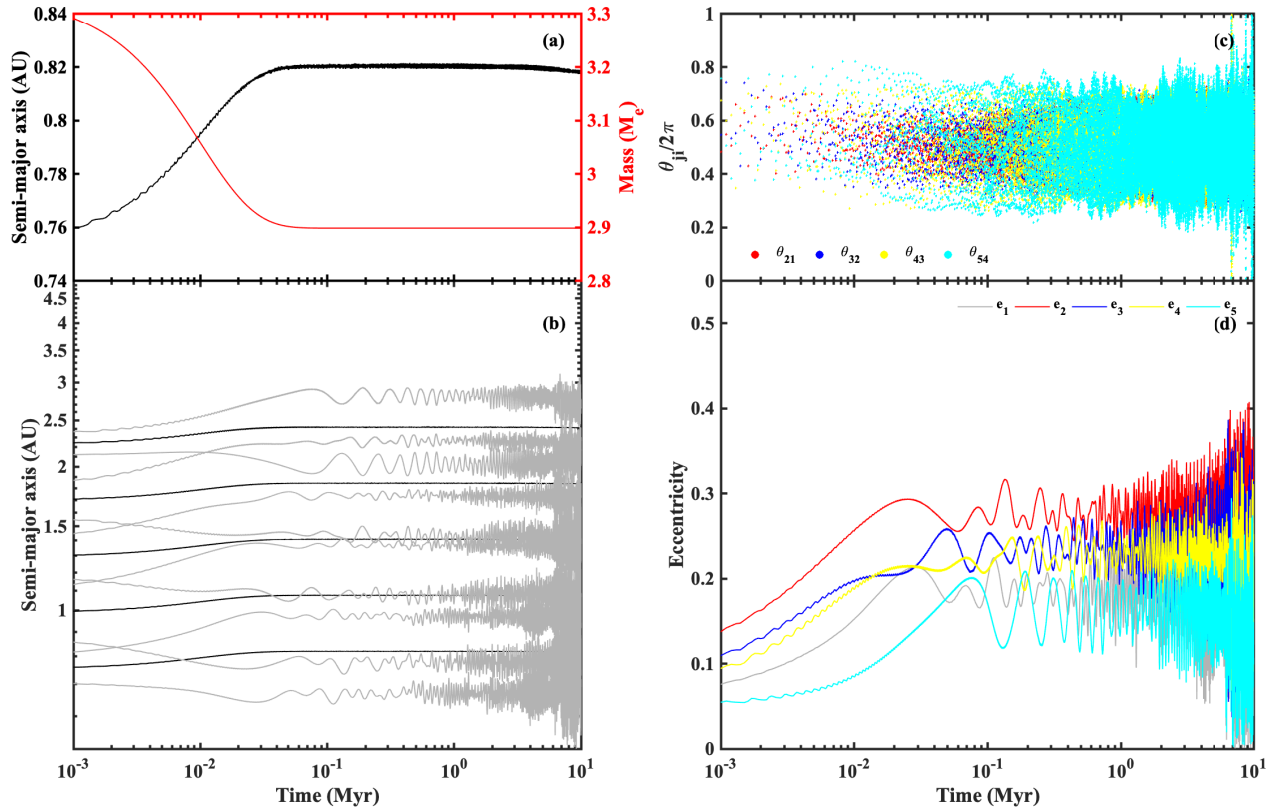}
 \caption{The evolution of the system with five planets, A=0.15 and $\tau_m=10^4$ yrs. Panel (a) shows the evolution of the innermost planet. In this case, the innermost planet will lose about 0.5 Earth-mass which is about 15\% of the initial mass. Panel (b) shows the evolution of a, a(1-e), and a(1+e) of all the planets. At the last stage of the simulation, the planet pairs in the system tend to escape from the MMRs. Panel (c) shows the evolution of the resonant angles. The resonant angles tend to be librate in a large range at the end of simulation. Panel (d) shows the evolution of eccentricities of planets in the system. The meanings of different colors are the same as in Figure \ref{case1}}.
 \label{marginal5}
 \end{center}
\end{figure*}

\begin{figure*}[htp]
\begin{center}
  \epsscale{1.1}
  \plotone{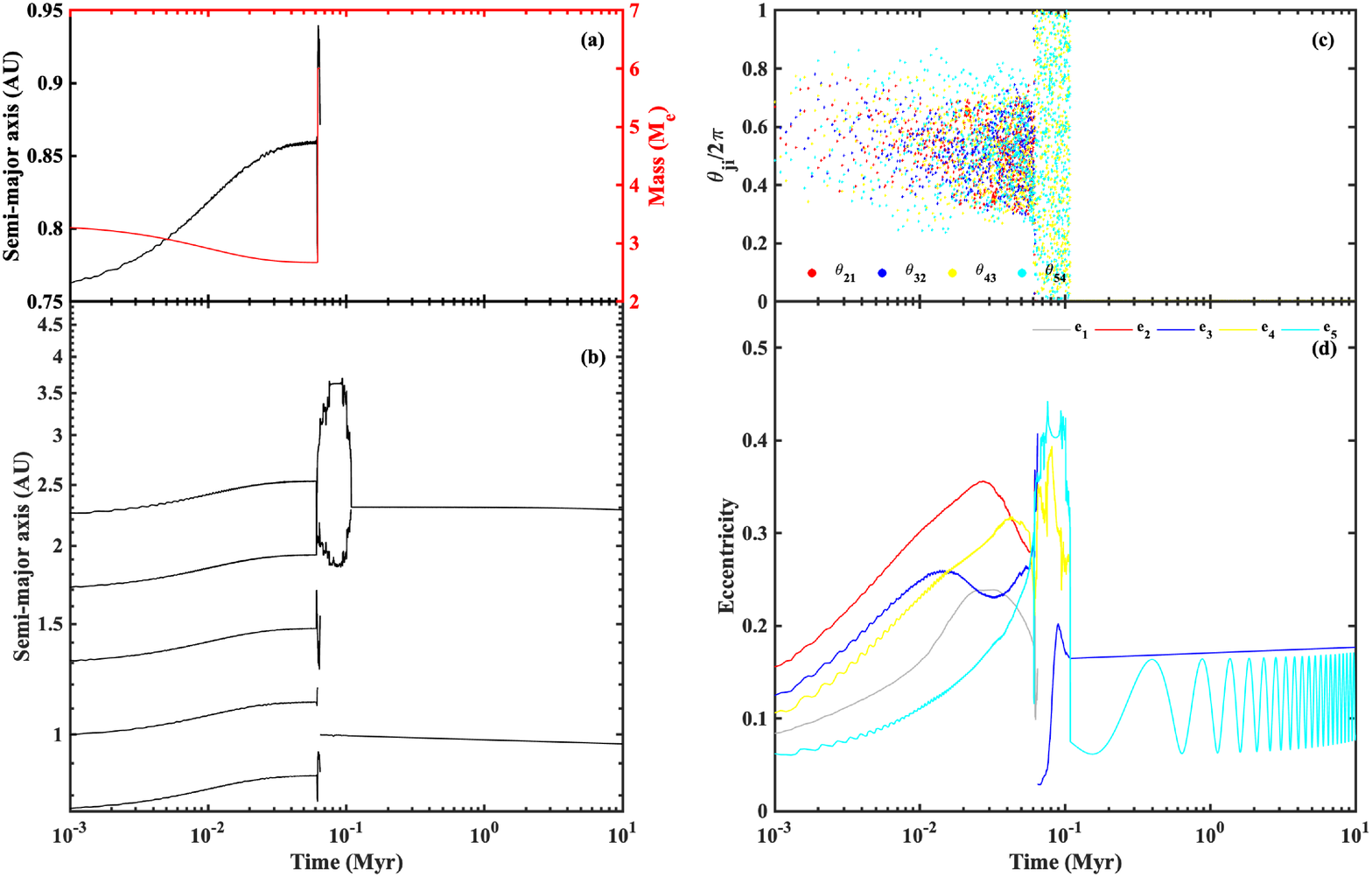}
 \caption{The evolution of the system with five planets, A=0.25 and $\tau_m=10^4$ yrs. The system become unstable at about 0.06 Myrs. Panel (a) shows the evolution of the innermost planet. After the system become unstable, the inner three planets collide into one larger planet which is about 9.33 $M_\oplus$. And then planet 4 and 5 collide into a planet with 6.67 $M_\oplus$. Panel (b) shows the evolution of a, a(1-e), and a(1+e) of all the planets.Panel (c) shows the evolution of the resonant angles. Panel (d) shows the evolution of eccentricities of planets in the system. The meanings of different colors are the same as in Figure \ref{case1}}.
 \label{unstable5}
 \end{center}
\end{figure*}

We generalize these calculations for systems which contain diverse
MMR chains.  These configurations are formed in disks with a range of
$\tau_{\rm dep} (=0.1, 1/3, 1,$ and 10 Myr) for the same value
of $f_g=0.2$ (models G3-3, G3-4, G3-5 and G3-6 in Table \ref{run1}).
Although their libration amplitudes for the adjacent resonant
pairs differ considerably, their evolution subsequent to the
incremental changes is very similar to that shown in Figure \ref{ct}.
For Model G3-3, the libration amplitude is relatively large (Fig.
\ref{am}) such that very small changes in the planets' spacing can
lead to the breaking of the MMRs.  Subsequent secular interaction
between these
multiple planets leads them to orbital crossing on time scales less
than 1 Myr.  With a relatively large $\tau_{\rm dep}$, Model G3-6
produces a more closely packed system.  Its small libration amplitudes
suggest the system contains deep and stable MMR chains.  Nevertheless,
relative small incremental changes in their spacing would shorten
the crossing time to less than 1 Myr. These results imply that
although multiple MMRs can be established in compact planetary systems
while they are embedded in their natal disks, the extent of these
islands of stability becomes narrowly confined after the depletion of the
disk gas, 10\% compression ($f_k\le 0.9$) of its original stable configuration can lead to quick orbital crossing, the crossing time  shorter than $10^4$ yrs.

\subsection{Preservation of MMRs during an extended epoch of mass-loss}
\label{sec:mmrmassloss}
In \S\ref{sec:origin} we show that as a consequence of tidal
interaction with their natal disks and type I migration, planets in multiple
planetary systems with MMR's robustly emerge. In disks which deplete
rapidly, these systems have large libration amplitude whereas in
more massive disks, they tend to be more compact.
Under most circumstances, their MMR configuration can be preserved for
at least 10 Myrs. But their islands of stability are surrounded by nearby
unstable regions where planets' mutual perturbation can trigger
dynamical instability and orbit crossing within a few Myrs (see
\S\ref{sec:island}). Small compressive or expansive impulsive adjustments
in the distance between the adjacent resonant pairs are more prone to
dynamical instability than those between the spaciously separated
multiple planetary systems (see \S\ref{sec:orbitcrossing}).

We now consider the stability and dynamical evolution of these
systems subjected to individual planet's atmospheric mass-loss
and orbital adjustment (see \S\ref{sec:massloss}). Because of their
confined islands of stability, the destiny of multiple planetary
systems with MMRs may provide more stringent constraints on amount of
mass-loss induced by the photo-evaporation of their atmosphere.

This process is expected to be protracted and persistent.
In order to simulate such process, we adopt the asymptotic
(at 10 Myrs) kinematic structure of an emerging multiple planetary
system (from model G2) as the initial conditions for later
dynamical evolution. Six typical cases are shown in Table \ref{6case}. The parameters are adopted
for the fractional amount ($A$) and time scale ($\tau_m$) of
mass-loss from the innermost planet.  In all six
cases, conservative mass-loss is applied to the same initial orbital
elements of the nine planets (Table \ref{run2}).

\begin{table}[htp]
\centering
 \caption{The parameters of mass-loss in the cases in section 3.2. A is the mass-loss fraction, $\tau_m$ is the mass-loss timescale, $\bar{\theta}/2 \pi$ is the average libration amplitude of all the planet pairs, and $\bar{e}$ is the average eccentricity of all planets in the system.}
\begin{tabular*}{8cm}{@{\extracolsep{\fill}}ccclllll}
\tableline
Case&A&$\tau_m$ (yr)& $\bar{\theta}/2 \pi$& $\bar{e}$\\
\tableline
1&0.05&$10^6$&0.029&0.147\\
2&0.05&$10^5$&0.029&0.150\\
3&0.05&$10^4$&0.27&0.179\\
4&0.1&$10^6$&0.17&0.195\\
5&0.1&$10^5$&0.18&0.230\\
6&0.1&$10^4$&0.36&-\\

\tableline
\label{6case}
\end{tabular*}
\end{table}

In case 1-3, we consider the innermost planet lost 5\% of its mass
($A=0.05$) on timescales ranging from $10^6-10^4$ yrs. Figures \ref{case1}
and \ref{case3} show the evolution of Case 1 and 3, respectively. Case 1
remains stable for 10 Myrs despite the large eccentricity (0.24) for
planet 2.  The resonant angles of all the MMR pairs remain
close to $\pi$. Their small average libration amplitudes (Figure
\ref{case1}) suggest that this system is locked into a deep resonant
chain.  The mass-loss time scale is sufficiently long to enable the planets
to adiabatically adjust their orbits and to share among themselves the angular
momentum change due to the mass-loss from the innermost planet. Their
MMR configuration is maintained.  Although the magnitude of $\tau_m$
in Case 2 is smaller than that in Case 1, the kinematic structure
of the system is also maintained for more than $10$ Myrs. In Case 3,
(with $\tau_{\rm dep} =10^4$ yrs) the planets' eccentricities $e$ and libration amplitude $\theta$ of the resonant angles
between adjacent MMR pairs grow substantially (Figure \ref{case3}).
The planets' average eccentricities $\bar{e}$ and average libration
amplitude $\bar{\theta}$ in Case 3 are much larger than those in
Case 1 (Table \ref{6case}). Although this system remains marginally
stable after 10 Myrs, chaotic relaxation is likely to trigger dynamical
instability and orbit crossing over a longer time scale.

In case 4-6, the fractional mass-loss is set to be 10\% (i.e.
$A=0.1$, higher than that in case 1-3) on time scales $\tau_m=10^6-10^4$ yrs.
The growth of planets' eccentricities and resonant angles between
adjacent MMR pairs in Case 4 and 5, shown in Table \ref{6case}, are similar to that in Case 3. In Case 6,
$\tau_m = 10^4$ yrs  is comparable to the libration
time scales for the resonant angles between the adjacent pairs
(in the range of a few $10^4$ yrs).  The rapid change of the innermost
planet's orbit can no longer be accommodated by adiabatic adjustment
of other planets such that the libration amplitude of resonant angles
($\bar{\theta}$) grow to $\pm \pi$ for some relevant pairs to break
their MMRs \citep{Liu15}, as shown in Figure \ref{case6}. This transition from libration to circulation
is followed by orbit crossing.  Later close encounters excite large
eccentricities.  Thus, with a relatively small ($\sim 5\%- 10\%$) fractional
mass-loss, compact systems with multiple MMRs have a tendency to become
unstable and undergo orbit crossing, especially for the system with higher mass-loss fraction or shorter mass-loss timescales.

For the systems with nine planets around the central star, the system tends to be destroyed by an orbital crossing event. However, if the number of planets in the system decreases, the results change. Up to now, the observed multiple planetary systems with more than three planets in the system are mainly focused on no more than five. Herein, we test the influence of mass-loss process in the system with three or five planets. The initial configurations are chosen from the innermost part of the results of G2 as shown in Table \ref{run2}. Figure \ref{marginal5} and \ref{unstable5} show two typical results. The trends of these groups with fewer planets in the systems are similar to that in nine-planet systems. With shorter mass-loss timescale or higher mass-loss fraction, the orbital crossing events are easier to happen in the system. As shown in Figure \ref{marginal5}, when the fraction of mass-loss reaches 15\% of the innermost planet's total mass, the amplitude of the resonant angles increase with the evolution, planet pairs in the system try to escape from the MMRs which is similar to that happened in Case 3 as shown in Figure \ref{case3}. But the critical fraction of mass-loss ($A_c$) that leads to the orbital crossing varies. With the decreasing of the number of planets in the system, $A_c$ increases. If the system contains five planets, $A_c$ increases to 25\%, while when the innermost planet loses almost 35\% of its total mass, orbital crossing happened in three-planet system. Additionally, after the orbital crossing happened, planetary systems which contains less than five planets are able to survive. Figure \ref{unstable5} shows the results with $A=$0.25 and $\tau_m=10^4$ yrs in a five-planet system. After the orbital crossing, the inner three planets collide to be a larger one, and the outer two planets also collide and merge to a larger planet. At the end of the simulation, there are two super-Earths (with mass ratio $m_{outer}/m_{inner}=1.4$) formed in the system with widely separated configuration (with period ratio of $\sim$ 3.6), like the configuration of system Kepler-1090 which contains two widely separated planets \citep{Morton16, Vali22}, the ratio of planet radius of these two planets is $\sim$ 1.46 and the period ratio is $\sim$ 4.68. 

Therefore, there are no clear indications of significant orbital adjustment induced by atmospheric losses in observed closely packed multiple planetary systems. Planets in such systems probably emerged from their natal disks without much primary atmosphere. While for the widely separated planetary systems, the planets may preserve a certain amount of atmosphere once before they suffered from strong photo-evaporation rising from the central star.

\section{Summary and Discussions}

In this work, we investigate the dynamical evolution of multiple terrestrial
planetary systems under the influence of orbital migration and mass-loss of planets caused by photo-evaporation.

We showed that sparsely populated multiple systems
(not initially in MMRs) can maintain their stability despite some
modest fractional ($\sim 10\%$) mass-loss
especially if it proceeds on a relatively long time scale or high planetary mass or larger relative space between the nearby planets.  Nevertheless,
it is difficult for a system to preserve its stability if one or more
planets lose 20\% of their total mass. Therefore, for the systems with planets not in MMRs, the stability of the system determined by the intensity of mass-loss process on the planets, the lowest mass of the planets, and the distance between the adjacent planets.

We adopt the assumption that super-Earths undergo
extensive type I migration while they are embedded in their natal
disks.  They emerge with diverse kinematic properties which are
determined by the surface density distribution and the depletion rate
of the gas disk.  Some of these systems are spatially extended while
others contain MMRs. The depletion timescale of gas disk will determine whether the system can
maintain stability after they are captured into MMRs. According to
the observation data, the depletion timescale of gas disk is few
million years \citep{hai01, WC11}.  Based on our results, the planetary system can maintain long-lived stability for $\tau_{dep}>1$ Myr which results in deep MMRs of planet pairs. The surface density of gas disk determines how compact the final configuration which also related to the stability of the systems. In general, such systems are dynamically stable
with relative small eccentricities and libration amplitudes.

Another factor influence the stability of the system during the orbital migration process is the location of the transition radius from the optically thick region to the optically thin region in a disk. If the planets formed in a disk 
with the transition radius very close to the central star, planets tend to undergo only inward migration and the system may be easy to be stable if there is a stopping mechanism. If the transition radius is far away from the central star, planets have large opportunity to form a compact configuration, just like the
configuration shown in Group 2 after both inward and outward migration. However, the island of stability of systems after the orbital migration process is very small. Tiny perturbations will destroy the system. Densely packed systems with resonant chains can break their mean motion resonances by sufficiently
large ($\geq 10 \%$) impulsive semi-major axis changes or persistent
atmosphere loss by one or more of the planets.

However, if the system formed around a star with strong
irradiation, planets which are in a MMR configuration may lose significant fractions of their mass. Such a process may destroy the stable configuration
they have already formed. With the effect of mass-loss on the innermost planet, the final configuration is related to the timescale competition between the mass-loss process and the libration of the resonant angles. When the libration timescale is larger than the mass-loss timescale, the amplitude of the resonance angles become large enough to break the resonance making orbital crossing happened in the system. With the timescale of mass-loss less than $2\times 10^4$ years, planet pairs may escape from MMR. The system can be destroyed if the innermost planet lost $\sim 10-20 \%$ of its total mass for the system with nine planets which is consistent with the results obtained from the mass-loss process in all first-order resonances multiple planetary systems \citep{Matsumoto20}. The system, however, can survive in a widely separated configuration if they are formed from the system with less than five embryos initially.

Based on our work, the destiny of the closely packed multiple planetary systems can be classified into three possible types. (I) The system is far away from the central star with MMRs configuration. Multiple planets in such systems undergo orbital migration which make them in or near MMRs and find their stable configuration finally. (II)The innermost planet is close to the central star and formed with little atmosphere. The system can keep in stable and chain resonances configuration after little mass-loss ( less than $\sim$5-10\% of its total mass). We find that systems similar in configuration to TRAPPIST-1 \citep{Gi17, Tam17} or K2-138 \citep{Lopez2019} can be stable for more than $10^7$ yrs if they are formed through orbital migration and without primary atmospheres. (III) The innermost planet is close to the central star and formed with a certain atmosphere. A stable configuration can be destroyed, leaving a single planet system or system with wide separated planets. From TESS mission results, there are more than 20 multiple planetary system candidates with at least three planets in the system have been observed as shown in Figure \ref{multi}. 35/59 planet pairs are in near first order MMRs, among them 20/59 are near 2:1 MMRs as labelled in red dots and circles, 17/59 are near 3:2 MMRs labelled in green dots and circles, and 3/59 are near 4:3 MMRs. A large proportion of systems contain MMRs configurations. Such systems may be stable for longer than $10^7$ yr, if they are formed through orbital migration, and the innermost planets have a very small amount of atmosphere, such as the systems TOI-1136 \citep{Dai21}, TOI-1749, TOI-175 \citep{Fukui21}, and TOI-178 \citep{Leleu2021}, they can be classified as type II. The systems with wide separated planets not in MMRs, like TOI-880 \citep{Sun22}, TOI-451 \citep{Newton21}, TOI-4010 \citep{Kunimoto22}, TOI-431 \citep{Osborn21}, and TOI-561 \citep{Lacedelli20} may have gone through a similar formation process with type III planetary systems.

In this work, we mainly focus on the influence of mass-loss process on the dynamical evolution and stability of multiple planetary systems. More than 800 multiple planetary system have been confirmed including 67 four-planet systems, 27 five-planet systems, 10 six-planet systems, 1 seven-planet system and 1 eight-planet system. For the systems with five or more planets in closely packed configurations, the innermost planet must not lose too much atmosphere, less than 10\% of its total planetary mass, to keep the system stable. Planets in multiple planetary systems undergoing rapid photoevaporation, especially those that lose a large fraction of their mass, are prone to orbital crossing which may destroy the system or leave fewer planets in widely separated configurations after fierce collisions and merger processes.  According to our results, long-term stable multiple planetary systems which formed through orbital migration from the outer disk favor a slower mass-loss process as concluded in the section \S\ref{sec:planetmmrs}. If the planets in multiple planetary system sustain mass-loss process for shorter than $\sim~2 \times 10^4$ yrs, the system will be faced with the challenge of being unstable or being destroyed. Therefore, for planets undergoing a core-powered mass-loss process, which acts on Gyrs instead of Myrs \citep{Ginzburg2018, Gupta2019}, the multiple planetary systems may remain in stable configurations for more than $10^7$ yrs. On the other hand, planets formed in-situ with minimal migration are likely to go through a "boil-off" process, with rapid atmospheric escape on timescales $\sim~10^5$ yrs \citep{Ikoma2012, Owen2016, Ginzburg2016}. Such multiple planetary systems are likely to be unstable, leaving fewer planets in widely-separated configurations. If, however, the planets lose less than $\sim$ 5\% of their total mass, the system may remain stable, even with a "boil-off" phase.

\begin{acknowledgments}
We thank the referee for a thorough and constructive report that significantly improved the manuscript. This work is supported by the B-type Strategic Priority Program of the Chinese Academy of Sciences, Grant No. XDB41000000, National Natural
Science Foundation of China (Grants No. 12033010,
11633009, 12111530175), the Natural Science Foundation of Jiangsu
Province (Grant No. BK20221563), the Strategic
Priority Research Program on Space Science of the Chinese Academy of Science (Grant No.
XDA15020800), the China Manned Space Project with No. CMS-CSST-2021-B09, the Foundation of Minor
Planets of Purple Mountain Observatory, and Youth Innovation
Promotion Association.
\end{acknowledgments}


\end{CJK*}
\end{document}